\documentclass[%
 reprint,
superscriptaddress,
 amsmath,amssymb,
 aps,
]{revtex4-2}

\usepackage{graphicx}
\usepackage{dcolumn}
\usepackage{bm}

\usepackage{siunitx}
\usepackage{caption}
\usepackage{subcaption}

\begin{document}

\preprint{APS/123-QED}

\title{CLIC Wake Field Monitor as a detuned Cavity Beam Position Monitor:\\
       Explanation of center offset between TE and TM channels in the TD26 structure}

\author{Kyrre Ness Sjobak}\email{k.n.sjobak@fys.uio.no}
\affiliation{Department of Physics, University of Oslo, 0316 Oslo, Norway}
\affiliation{CERN, CH-1211 Geneva 23, Switzerland}

\author{Hikmet Bursali}
\affiliation{CERN, CH-1211 Geneva 23, Switzerland}
\affiliation{Department of SBAI, Sapienza University of Rome, Italy}

\author{Antonio Gillardi}
\affiliation{CERN, CH-1211 Geneva 23, Switzerland}
\affiliation{University of Naples Federico II, DIETI - IMPALab, Napoli, Italy}

\author{Reidar Lillestøl}
\affiliation{Department of Physics, University of Oslo, 0316 Oslo, Norway}

\author{Wilfrid Farabolini}
\affiliation{CERN, CH-1211 Geneva 23, Switzerland}

\author{Steffen Doebert}
\affiliation{CERN, CH-1211 Geneva 23, Switzerland}

\author{Erik Adli}
\affiliation{Department of Physics, University of Oslo, 0316 Oslo, Norway}

\author{Nuria Catalan Lasheras}
\affiliation{CERN, CH-1211 Geneva 23, Switzerland}

\author{Roberto Corsini}
\affiliation{CERN, CH-1211 Geneva 23, Switzerland}

\date{\today}

\begin{abstract}
The Wake Field Monitor (WFM) system installed on the CLIC prototype accelerating structure in CERN Linear Accelerator for Research (CLEAR) has two channels for each horizontal/vertical plane, operating at different frequencies.
When moving the beam relative to the aperture of the structure, a disagreement is observed between the center position of the structure as measured with the two channels in each plane.
This is a challenge for the planned use of WFMs in the Compact Linear Collider (CLIC), where they will be used to measure the center offset between the accelerating structures and the beam.
Through a mixture of simulations and measurements, we have discovered a potential mechanism for this, which is discussed along with implications for improving position resolution near the structure center, and the possibility determination of the sign of the beam offset.
\end{abstract}

\maketitle

\begin{figure}
    \centering
    \includegraphics[width=\linewidth]{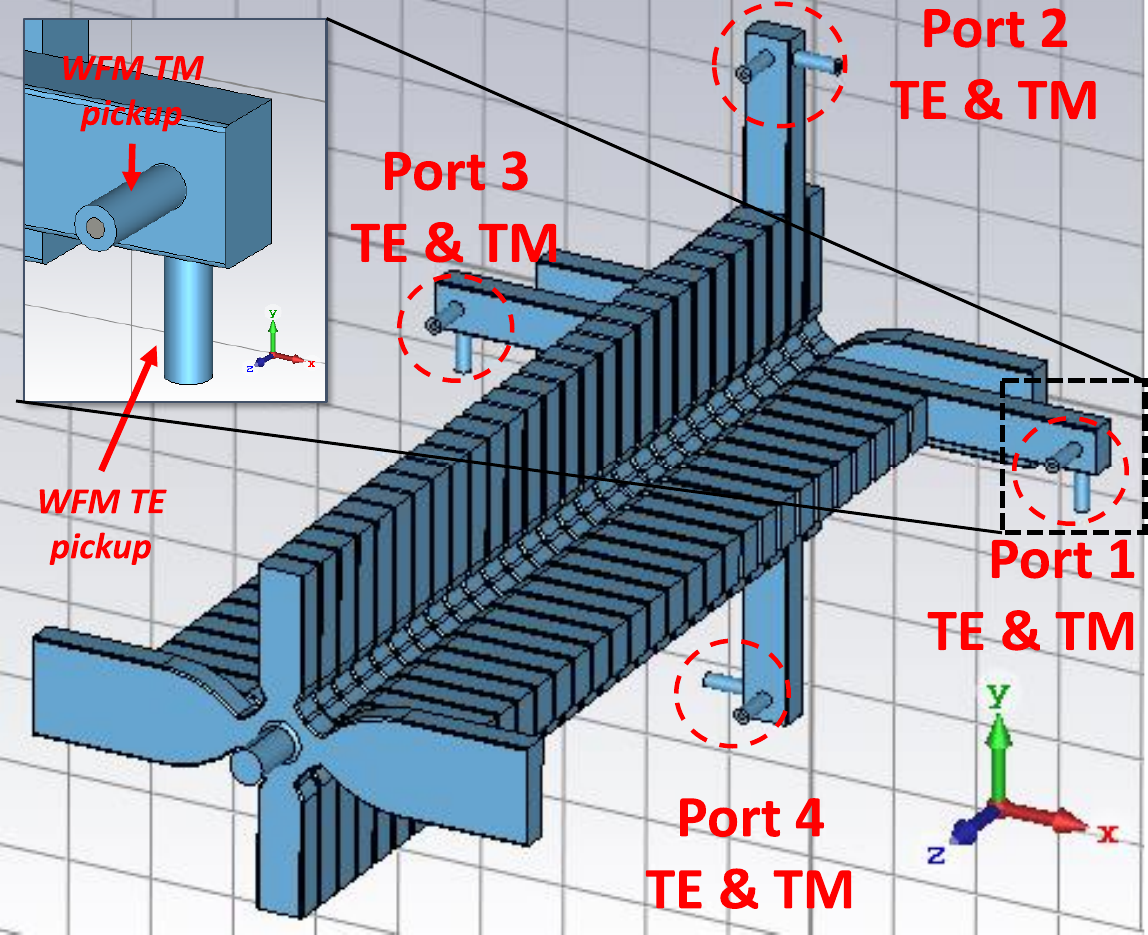}
    \caption{3D model of the TD26 structure mounted at CLEAR, with antenna locations highlighted. TE and TM antenna positions are described in the inset, with TM antennas generally being on the long side of each DWG, and the TE antennas on the short side.}
    \label{fig:structure-antennas}
\end{figure}

The Wake Field Monitors (WFMs)~\cite{peauger_wakefield_2011} are a crucial system in order to reach the luminosity goals of the Compact Linear Collider (CLIC)~\cite{clic_collaboration_compact_2018,clic_collaboration_updated_2016,clic_collaboration_multi-tev_2012,kraljevic_beam-based_2019}.
These devices monitor the beam position relatively to center of the 12~GHz high gradient accelerating structures, with one device mounted near the middle of each super-structure.
This makes it possible to correct the position of the structures relatively to the beam, which minimizes the transverse wake-field effect on the beam.
This allows efficient use of the particles at the interaction point and thus maximizes the luminosity.

In order to verify the operation, resolution, and accuracy of the CLIC WFMs, one such accelerating structure~\cite{gilardi_measurements_2021,lillestol_reidar_lunde_clic_2017,lillestol_status_2016,lillestol_wakefield_2015,quirante_califes_2014,farabolini_recent_2014} is mounted in the beamline of the CERN Accelerator for Research (CLEAR)~\cite{gamba_clear_2017,sjobak_status_2019}.
This development system has two channels for each plane, working at different frequencies (18 and 24~GHz).
Both of these channels are expected to give minimum signal when the beam is in the electrical center of the accelerating structure.
However, this was not observed experimentally: A discrepancy in the position of the signal minimum on the order of \SI{100}{\micro m} was observed.
Furthermore, this discrepancy is not stable, but varying with beam conditions.

Since we are interested in knowing the position of the beam relatively to the electrical center of the accelerating structure (i.e.\ where there is no kick~\cite{gilardi_measurements_2021}), this discrepancy must be understood.

Wake-field monitors are used or have been tested for structure alignment and diagnostics at several facilities, including X-band linearizers at the SwissFEL~\cite{dehler_x-band_2009,dehler_alignment_2009,dehler_wake_2013,dehler_front_2014} and tests of damped detuned structure (DDS) accelerating structures for the Next Linear Collider~\cite{seidel_studies_1997,adolphsen_wakefield_1999,dobert_beam_2005}, L-band cavities for TESLA~\cite{molloy_high_2007,frisch_high_2006,baboi_preliminary_2004}, and S-band cavities for the CLIC Test Facility drive beam~\cite{prochnow_measurement_2003}.
These structure were not as highly damped as the structure studied in this paper, meaning that the wake field has narrower modes and less overlapping frequency spectrum than the structure used in this work, and consequently also longer decay times.
In at least some of these structures, the modes are localized to single cells, separated by frequency~\cite{dobert_beam_2005}.

The RF properties of the WFM system for damped CLIC structures have also been studied without beam using the TD24 prototype accelerating structure~\cite{munoz_electromagnetic_2015,munoz_pre-alignment_2016,munoz_pre-alignment_2017} and with beam using both TD24~\cite{quirante_califes_2014,farabolini_recent_2014} and TD26~\cite{lillestol_wakefield_2015, lillestol_status_2016, lillestol_reidar_lunde_clic_2017}.
The results from the beam tests of the TD24 structure were encouraging, indicating that precise position measurement is possible.
However, the geometry of this structure and the installation of the Wake Field Monitors was different from the TD26 structure planned for CLIC, prompting the experiments reported in this paper.

\begin{figure*}[tb]
    \centering
    \includegraphics[width=\linewidth]{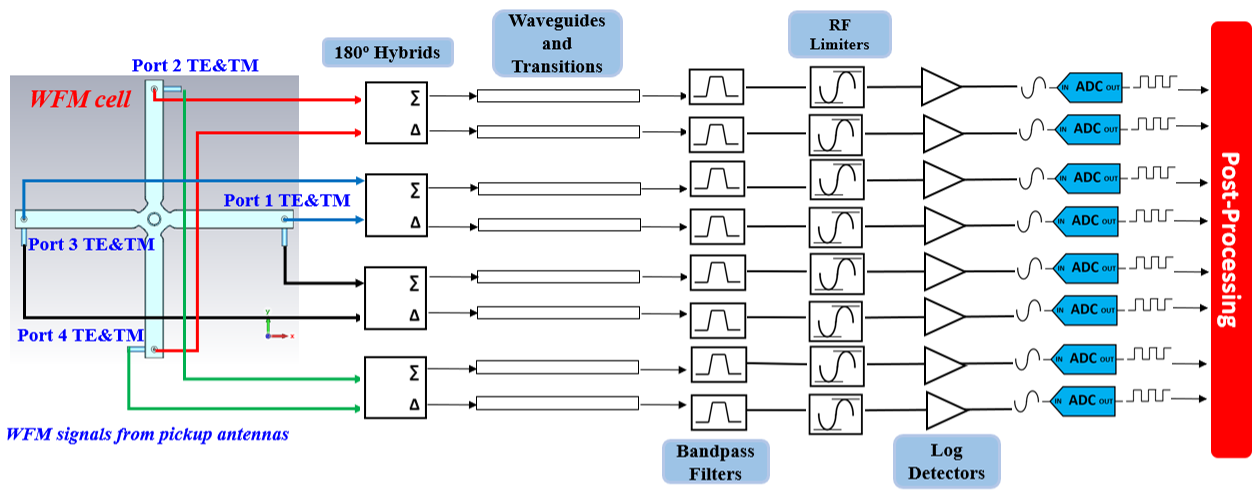}
    \caption{Schematic for the RF front-end of the CLEAR WFM system, showing how the digitized signals for post-processing are normally acquired from the accelerating structure.}
    \label{fig:wiring-schematics}
\end{figure*}

\section{Experimental setup}

To reduce the long range wake potential, each cell of the CLIC 12~GHz high-gradient accelerating structures are equipped with four higher order mode damping waveguides (DWGs), which are designed to have a cutoff frequency of 13.63 and 22.50~GHz in the two first TE polarizations~\cite{sjobak_design_2014}.
These are terminated with silicon carbide loads~\cite{de_michele_analysis_2012,grudiev_design_2010}, reducing the Q-factor of the beam-excited higher order modes (HOMs) to values of around~10.
In the accelerating structure installed at CLEAR, as shown in Fig.~\ref{fig:structure-antennas} the DWGs of first regular cell of the 2nd structure is extended, and two pickup antennas are placed on each side of the rectangular waveguide; the accelerating structure thus has a total of 8 antennas.

\subsection{RF front-end}
\label{sec:setup:RF}

The signals from the wake field monitoring system are expected to be proportional to the beam offset from the electrical center of the structure.
Furthermore, the signals from the TM-like modes are expected to propagate into one pair of DWGs with a \SI{180}{\degree} phase shift, whereas the TE-like modes are expected to produce in-phase signals.
For this reason, the signals are combined in pairs using RF hybrids as illustrated in Fig.~\ref{fig:wiring-schematics}, producing the sum ($\Sigma$) and difference ($\Delta$) of the antenna signals.
This rejects common-mode noise due to the presence of other modes, increasing the signal-to-noise ratio.

\begin{figure}
    \centering
    \includegraphics[width=\linewidth]{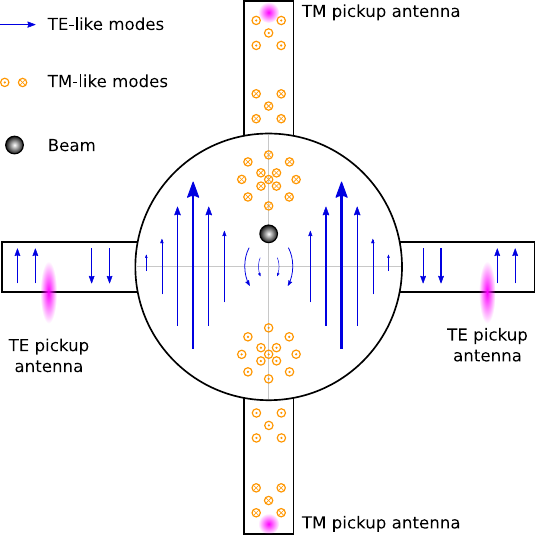}
    \caption{Illustration of the electric field distribution of beam-excited higher order modes due to an electron beam displaced vertically relative to center of the structure. Circles with cross represent vectors going into the page, circles with dot are vectors coming out of the page. Also indicated are the 4 pickup antennaes relevant for vertical beam displacement.}
    \label{fig:HOM-principle}
\end{figure}

With 8 antennas, the pairs with antennas on the wide side of the DWG (i.e.\ the antennae parallel to the direction of beam propagation) are called TM-mode pickups, while those with antennae on the short side of the DWG are TE-mode pickups.
These are named such because they are mainly sensitive to TM- or TE-like HOMs in the structure, due to the general transverse or longitudinal direction of the electric field in such modes.
As illustrated in Fig.~\ref{fig:HOM-principle}, the TM-like modes are primarily radiating out into the DWGs in the plane of beam displacement, while the TE-like modes are radiating out in the plane normal to the beam displacement.
Note that as seen in Fig.~\ref{fig:structure-antennas}, the TE-mode pickup antennae on the vertical pair of arms are located on opposite sides, giving an extra phase shift of \SI{180}{\degree} between the output signals from the two antennae.
This causes a horizontal beam offset to be visible in the $\Delta$ output for both the TE and TM signals.
For a vertical beam offset, the signals appear in the TM $\Delta$ and the TE $\Sigma$ channels.

\begin{figure}
    \centering
    \includegraphics[width=\linewidth]{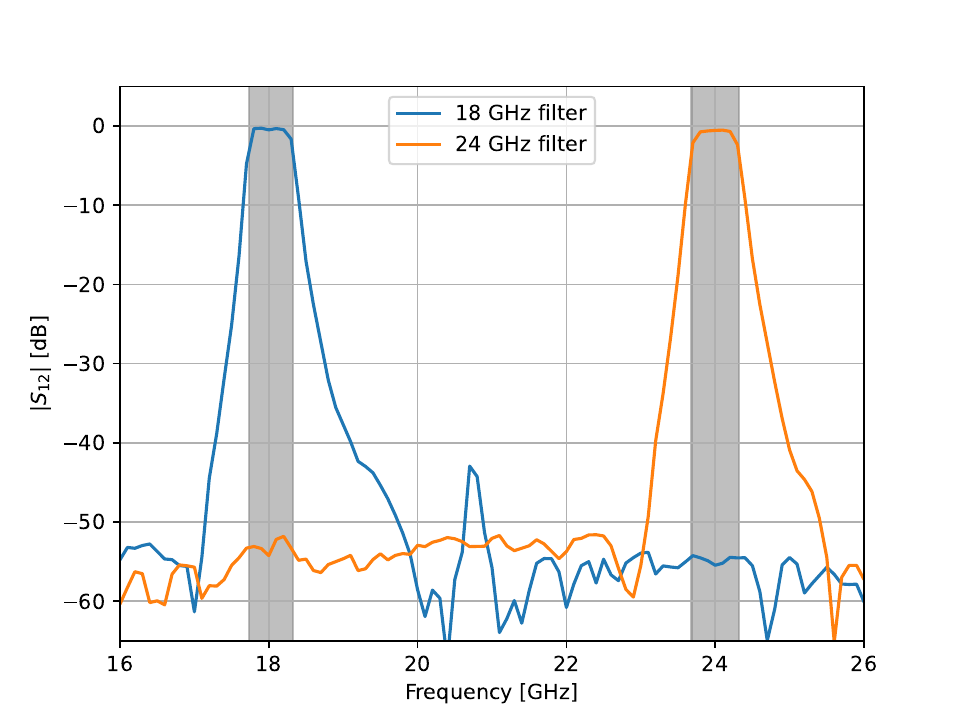}
    \caption{Measured attenuation as a function of frequency for the 18- and 24~GHz band pass filters used to isolate the TM- and TE-signals. The gray background indicate the FWHM in power.}
    \label{fig:filters}
\end{figure}

\begin{figure}
    \centering
        \includegraphics[width=\linewidth]{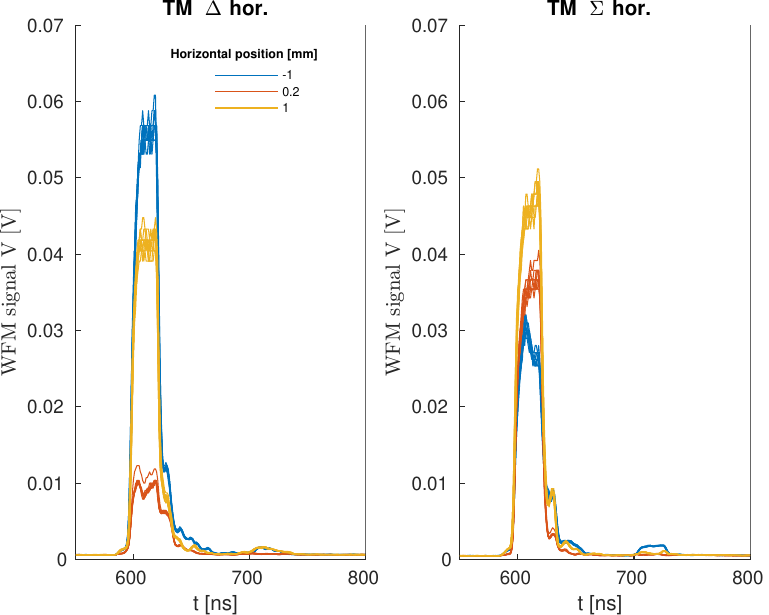}
        \includegraphics[width=\linewidth]{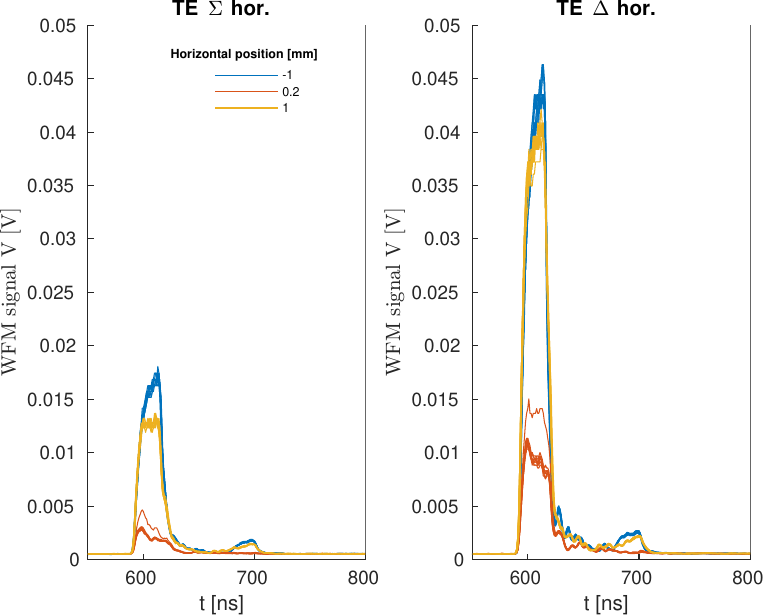}
    \caption{Typical linearized measured signals from WFM, using 30~bunches with a variable horizontal offset created with a pair of upstream kicker magnets. The bunch charge was 257~pC, with a bunch length 3.51~ps. Only the region with signal, between $t_1=550~\mathrm{ns}$ and $t_2=800~\mathrm{ns}$ relative to the beam synchronous trigger signal, is shown.}
    \label{fig:signals}
    %
\end{figure}

After the RF hybrids, the $\Sigma$ and $\Delta$ signals are passed through RF band-pass filters with central frequency of 18 (for TM) and 24~GHz (for TE), as shown in Fig.~\ref{fig:filters}.
Both filter types have a FWHM in power of 0.6~GHz.

The filters are followed by analog log-detectors (Analog Devices HMC662LP3E), which have a frequency range of 8-30~GHz and a dynamic range of 54~dB, and produce an output voltage $V_\text{log}(t)$ that is proportional to the input power in dBm.
These log-detectors are protected from overpowering by +10~dBm RF power limiters, and the output of the log-detectors are sampled by \SI{1}{\giga sample \per s} 8-bit Acquiris DC270 acquisition cards.

After sampling, the signals are linearized using a calibration polynomial that has been measured for each channel and for 18- and 24-GHz, converting it to a power level in dBm:
\begin{equation}
    P_\text{dBm}(t) = c_1 V_\text{log}(t) + c_2
    \label{eq:PowerDbmCallib}
\end{equation}
The calibration is valid for power levels between -45 and +5~dBm.
Assuming a single-frequency signal, the equivalent root-mean-square (RMS) voltage in a \SI{50}{\ohm} transmission line is given as
\begin{equation}
    V(t)~[V] = \sqrt{ \frac{10^{\frac{P_\text{dBm}(t)}{10}}}{1000} \SI{50}{[\ohm]} }\;.
    \label{eq:Voltage50Ohms}
\end{equation}
The cross-talk between the channels of the log-detector was also evaluated by monitoring the output of all channels while injecting power into each single channel.
In all cases it was found to be below -30~dB~\cite{gilardi_measurements_2021} relative to the power seen in the injected channel.

The typical shape of the linearized beam-generated signal detected by the log-detector for a TM- and TE-mode is shown in Fig.~\ref{fig:signals}.
In this signal, the pulse width depends on the train length, and the height on the bunch charge and beam position.
Some reflections are also visible after the main peak of the signals.

\subsection{Beamline equipment and accelerator setup}

\begin{figure*}
    \centering
    \includegraphics[width=\textwidth]{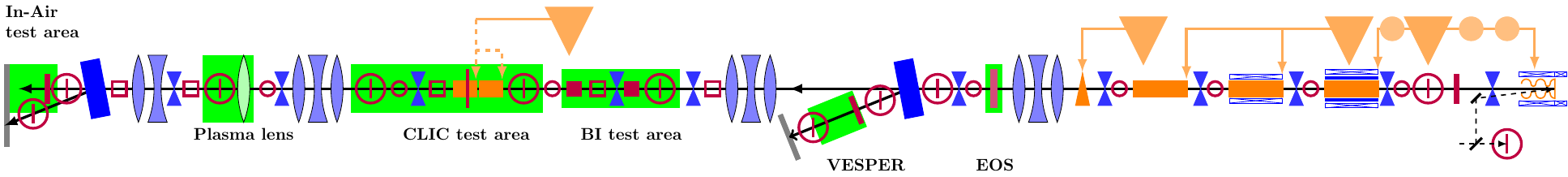}
    \caption{CLEAR beamline elements~\cite{sjobak_status_2019,sjobak_clear_2020} as it was configured during these experiments. The CLIC RF structure with WFMs is located in the ``CLIC test area'', and correctors (blue double triangles) used to scan the beam position are located just upstream in and around the ``BI test area''.}
    \label{fig:beamline}
\end{figure*}

As indicated in Fig.~\ref{fig:beamline}, the beam at CLEAR is generated by a photo-injector, enabling tuning of both the bunch charge and the number of bunches per accelerator pulse.
These bunches, which are generated at a frequency of 1.5~GHz, are then accelerated up to 200~MeV/c by three 3~GHz S-band accelerating structures.
The bunch length can be manipulated by changing the phase of the gun and/or the first accelerating structure.

After the accelerating section, there is a diagnostic section for measuring the bunch length~\cite{arpaia_enhancing_2020,sabato_measurement_2021}, beam energy and spread, and the transverse beam profile and phase space.
This is used when tuning the RF phases and trajectory in the linear accelerator.
Furthermore, the beamline is equipped with 3 integrating beam current monitors, measuring the total charge of the beam right out of the gun, in the energy measurement spectrometer (VESPER), and at the end of the beam-line.
The number of bunches is monitored on-line with a photodiode and a fast oscilloscope at the LASER; it can also be counted by gradually opening the pulse picker window and counting the number steps in the bunch charge as a function of window width.

Around the CLIC structure, we have various instruments for monitoring the beam profile and beam losses.
This is used to verify that the beam is correctly focused at the structure entrance, and for measuring the wake-field kick after the structure~\cite{gilardi_measurements_2021,arpaia_beambased_2019}.
An online optics model is used to optimize the beam envelope~\cite{sjobak_status_2019,sjobak_clear_2018}.

The structure itself is mounted on a transverse mover, making it possible to change the transverse position of it by remote control.
The position of the structure is monitored by remote-readout position gauges with \si{\micro\meter} accuracy~\cite{sjobak_clear_2021}.

Furthermore, the beam can be displaced relatively to the structure using the two corrector magnets between the structure and the quadrupole triplet in front of it.
This can both shift the beam trajectory through the structure parallel to the initial trajectory, and also change the angle of the beam trajectory.
The accuracy of the parallel shifts are better than 1\%~\cite{gilardi_measurements_2021}.

Note that when changing the beam charge at CLEAR, the beam trajectory out of the S-band LINAC generally changes as well.
While CLEAR has a system of BPMs and beam profile monitors, they are unfortunately not well enough characterized or aligned to accurately measure the absolute position of the beam across a range of beam conditions.
Because of this we are only able to measure relative position changes, e.g.\ between the TM and TE channel within the same scan, or within an uninterrupted sequence of scans with identical beam parameters.
This makes studies of the response as a function of the absolute position in the CLIC structure difficult, as the nearby BPMs are currently not able to achieve the required resolution.
However, studies of the response of the WFM can still be undertaken through relative measurements, as are reported in this paper.

\section{Measurement method}

\begin{figure}
    \centering
    \includegraphics[width=\linewidth]{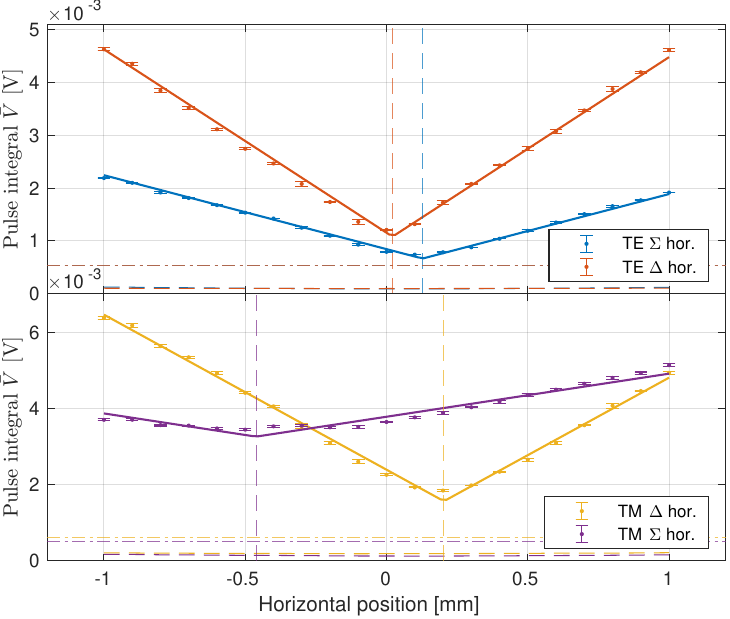}
    \caption{Pulse integral for TE and TM as a function of beam position relative to an arbitrary offset, fitted to Eq.~\eqref{eq:Vfit}, with the fitted centers $x_0$ indicated with vertical dashed lines. Same data as shown in Fig.~\ref{fig:signals}, with 10 shots per position. For each beam position, the mean of $\bar{V}$ is shown as the dot, and the estimated error of the mean as the errorbars. The dashed-dot line indicates the noise floor estimate, and the dashed line indicates cross-talk estimated using Eq.~\eqref{eq:Vcrosstalk}.}
    \label{fig:V-hybrid-standard}
\end{figure}

The position of the beam relatively to the CLIC structure was changed either by physically moving the structure transversely, or by use of the upstream kicker magnets.
The RF signals were recorded as described in Sec.~\ref{sec:setup:RF}, measuring the signals produced by a number of accelerator pulses for each relative beam position.
The signals, shown in Fig.~\ref{fig:signals}, are integrated within a pre-defined window $[t_1, t_2]$ that covers the entire voltage pulse, giving the pulse integral
\begin{equation}
    \bar{V} = \frac{\int_{t_1}^{t_2} V(t) \;\mathrm{d}t}{t_2-t_1}~.
    \label{eq:signalIntegration}
\end{equation}
This is proportional to the magnitude of the voltage seen by the log detector, and is independent of the signal phase.
The pulse integrals are then grouped for each relative beam position, and their mean and root-mean-square (RMS) are found.
This is then plotted as shown in Fig.~\ref{fig:V-hybrid-standard}, with the error bars representing the estimated error of the mean, $\mathrm{RMS}/\sqrt{N}$ where $N$ is the number of beam pulses at that position.

Also estimated is the noise floor, which is found by evaluating Eq.~\eqref{eq:signalIntegration} with $t_1$ and $t_2$ set so that the window is prior to the beam arrival.
Furthermore, for each bunch train, the cross-talk noise $\bar{V}^*_j$ to channel $j$ is estimated assuming incoherent signals, as
\begin{equation}
    \bar{V}^*_j = \sqrt{\sum_{\forall i \ne j}  \left( \bar{V}_i \, c_{i,j} \right)^2 }
    \label{eq:Vcrosstalk}
\end{equation}
where $c_{i,j}$ is the ratio of the measured voltage in channel $i$ and the injected signal voltage in channel $j$.
This ratio was measured during log detector calibration.

The ``V-function''
\begin{equation}
    V_\mathrm{fit} (x;A,x_0,V_0) = A \left|x-x_0 \right| + V_0
    \label{eq:Vfit}
\end{equation}
is finally fitted to the means of the grouped pulse integrals data using a weighted nonlinear least-squares fitting routine, where~$A$, $x_0$, and~$V_0$ are free parameters representing respectively the sensitivity, center position, and pedestal of the system.
This fit is also shown in Fig.~\ref{fig:V-hybrid-standard}, and the values of $x_0$ for each channel is shown in Tab.~\ref{tab:V-hybrid-standard}.
Generally, a strong response is seen in the plane where the beam is moved, and little response is visible in the other plane.

The error of this fit is estimated with Monte-Carlo sampling; 10'000 ``scans'' were generated by smearing the measured means of the pulse integrals at each position with a Gaussian distribution.
Each of these generated scans is then fitted, and the error estimates for are the RMS of the resulting fit parameters from the generated scans.

\begin{table}
    \centering
    \caption{Fitted center positions $x_0$~\si{\micro\meter} as defined in Eq.~\eqref{eq:Vfit}, with error estimate, for the data presented in Fig.~\ref{fig:V-hybrid-standard}.}
    \label{tab:V-hybrid-standard}
    \begin{tabular}{c c | c c}
        \hline
        \multicolumn{2}{c|}{TE data} & \multicolumn{2}{c}{TM data} \\
        \hline
        TE $\Sigma$: & $128 \pm 1$ & TM $\Delta$: & $203 \pm 1$ \\
        TE $\Delta$: & $21 \pm 1$ & TM $\Sigma$: & $-462 \pm 6$ \\
        \hline
    \end{tabular}
\end{table}

\section{Signal formation model}
\label{sec:signalformation}

As stated above, the goal of the WFMs are to measure the beam position relatively to the electrical center of the structure.
This enables the minimization of this offset by changing the relative position of the beam and the structure, reducing the transverse wake-field kicks on the beam and thus the emittance dilution.
This requires the electromagnetic center of the structure as measured by the WFMs to be well-defined relative to the kick center.
However, as seen in Fig.~\ref{fig:V-hybrid-standard} and Tab.~\ref{tab:V-hybrid-standard}, the minima ($x_0$) from the TE- and TM-signals do not coincide, which makes the comparison with the kick center difficult.

In order to interpret this data, we will consider a simple model for the signal formation, where the antennas are producing a single-frequency voltage which is linearly dependent on the beam offset $x$ relatively to the electrical center $c$:
\begin{equation}
    \tilde V(x,t) = A\exp(\mathbf{i} \omega t) (x-c)
    \label{eq:Vphasor-single}
\end{equation}
Here $\omega=2\pi f$ is the frequency of the signal, $A$ a complex constant representing the amplitude and phase of the signal, and $\mathbf{i}$ the imaginary unit. $\tilde V$ is the voltage signal in phasor form such that $V = \Re (\tilde V)$.
Since we are measuring the power after the filter, the detected signal is proportional to the magnitude of the (complex) voltage.
Assuming a different antenna gain, cable attenuation, and propagation delay on each side of the hybrid, the output signals from the hybrid is on the form
\begin{equation}
    \begin{split}
    \tilde V_\Sigma =& D_L e^{\mathbf{i}\theta_L} \tilde V_L(x,t) + D_R e^{\mathbf{i}\theta_R} \tilde V_R(x,t) \\
    \tilde V_\Delta =& D_L e^{\mathbf{i}\theta_L} \tilde V_L(x,t) - D_R e^{\mathbf{i}\theta_R} \tilde V_R(x,t) \;,
    \end{split}
    \label{eq:Vphasor-sumDelta}
\end{equation}
where $\tilde V_L(x,t)$ and $\tilde V_R(x,t)$ are the signals coming from each of the two arms, which are both assumed to be on the form $\tilde V(x,t)$ as defined in Eq.~\eqref{eq:Vphasor-single}.
In this, the constants $c$ and $A$ inside $\tilde V_R$ and $\tilde V_L$ are in the most general case assumed to be different and are labeled $c_L$, $c_R$, $A_R$, and $A_L$.
Furthermore $D_L$ and $D_R$ are the attenuation factors from the cable and antenna of each arm, and $\theta_L$ and $\theta_R$ are the phase shifts.

\begin{figure}
    \centering
    \begin{subfigure}{\linewidth}
    \includegraphics[width=\linewidth]{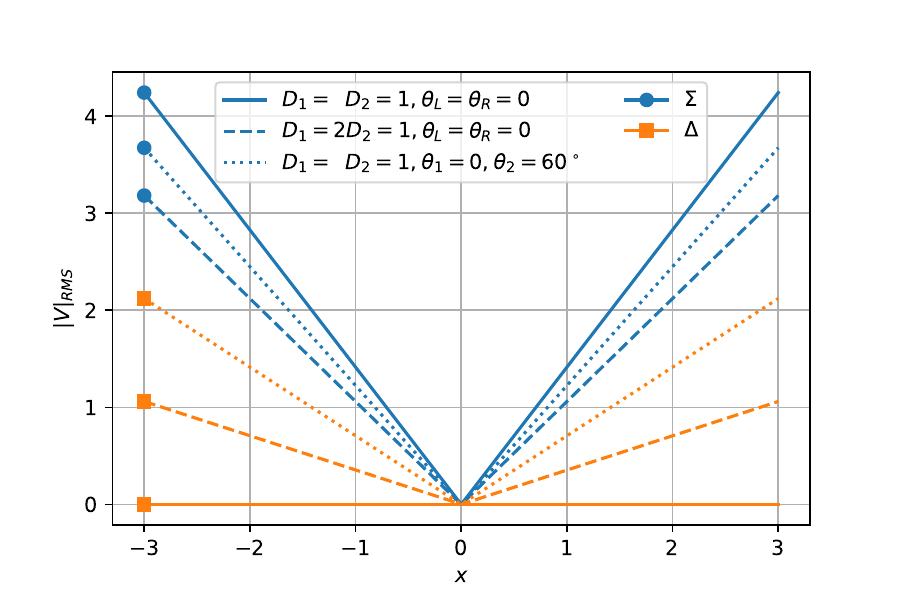}
    \caption{Single center $c=0$, as in Eq.~\eqref{eq:Vphasor-sumDelta:symmetric}.}
    \label{fig:V-simplemodel:centered}
    \end{subfigure}
    \begin{subfigure}{\linewidth}
    \includegraphics[width=\linewidth]{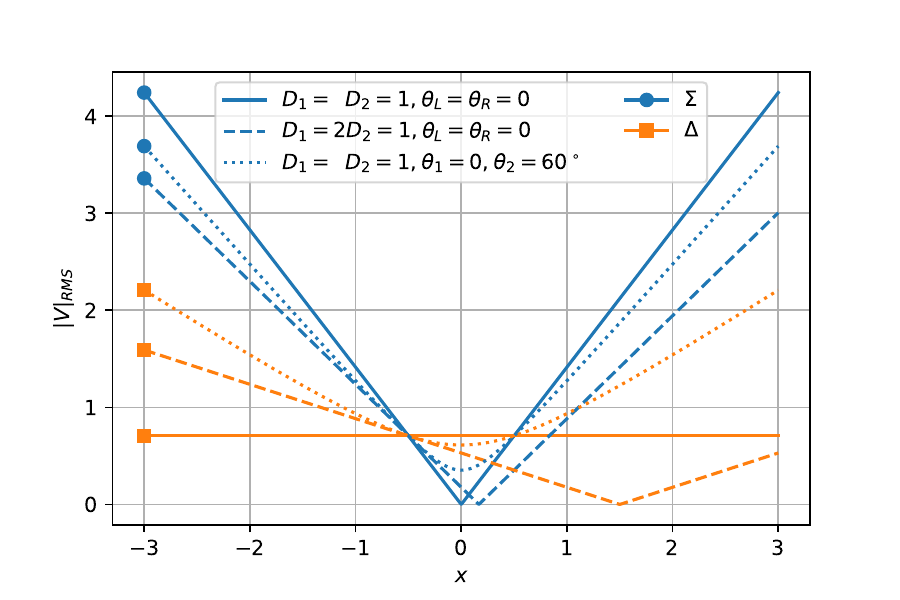}
    \caption{Different centers as in Eq.~\eqref{eq:Vphasor-sumDelta:shifted}, with $c'=0.5$.}
    \label{fig:V-simplemodel:shifted}
    \end{subfigure}
    \caption{Analytical single-frequency model of combined signal with various attenuation and phase imbalances. Vertical axis on plots is $|V|_{RMS} = |\tilde V / \exp(\mathbf{i} \omega t)|/\sqrt{2}$, which is equivalent to the WFM signal voltage amplitude $V$. Attenuation, phase, and center displacement parameters are chosen for demonstration.}
    \label{fig:V-simplemodel}
\end{figure}

In the case where $\tilde V_L(x,t) = \tilde V_R(x,t) = \tilde V(x,t)$ such that $c_L = c_R = c$ and $A_L=A_R=1$,
\begin{equation}
    \begin{split}
    \tilde V_\Sigma =& \left(D_L e^{\mathbf{i}\theta_L} + D_R e^{\mathbf{i}\theta_R} \right) \tilde V(x,t)\\
    \tilde V_\Delta =& \left(D_L e^{\mathbf{i}\theta_L} - D_R e^{\mathbf{i}\theta_R} \right) \tilde V(x,t) \;.
    \end{split}
    \label{eq:Vphasor-sumDelta:symmetric}
\end{equation}
Here the effect of $D_L \ne D_R$ or $\theta_L \ne \theta_R$ is to shift the RF power between the $\Sigma$ and $\Delta$ channels.
This produces a ``V-plot'' as described by Eq.~\eqref{eq:Vfit} in one or both of the $\Sigma$ and $\Delta$ channel.
The minimum amplitude will always be at $x=c$, as illustrated in Fig.~\ref{fig:V-simplemodel:centered}.

In the case where the two antennas see a different center $c_L$ and $c_R$, the minimum amplitude of the combined signals will depend on the phase shifts and attenuations.
Again assuming $A_L=A_R=1$, and that the displacement of the centers is symmetric such that $c_L = - c_R = c'$, Equation~\eqref{eq:Vphasor-sumDelta} becomes:
\begin{equation}
    \begin{split}
    & \tilde V_\Sigma = \exp(\mathbf{i}\omega t) \\
    &\left[(D_Le^{\mathbf{i}\theta_L} + D_Re^{\mathbf{i}\theta_L})x - (D_Le^{\mathbf{i}\theta_R} - D_Re^{\mathbf{i}\theta_R}) 
    c' \right] \\
    & \tilde V_\Delta = \exp(\mathbf{i}\omega t) \\
    &\left[(D_Le^{\mathbf{i}\theta_L} - D_Re^{\mathbf{i}\theta_L})x - (D_Le^{\mathbf{i}\theta_R} + D_Re^{\mathbf{i}\theta_R}) c' \right]
    \end{split}
    \label{eq:Vphasor-sumDelta:shifted}
\end{equation}
This is illustrated in Fig.~\ref{fig:V-simplemodel:shifted}.
We see that the effect $D_L \ne D_R$ is a shift of the center as observed in one channel, and the creation of a V in the other channel.
Furthermore, a phase shift in one channel creates a softening of the bottom of the V.
Finally, when $c' \ne 0$, the $\Sigma$ channel is lifted above 0 even when $D_L=D_R$ and $\theta_L = \theta_R$.

In a real installation, it is unlikely that the two signal paths are perfectly balanced, i.e.\ that $D_L=D_R$ and $\theta_L=\theta_R$.
Therefore, in the case where $c_L \ne c_R$, this produces an uncertainty of the electrical center as measured by the WFMs.
It is therefore necessary to check if this is the case for any of the signals produced by the WFM antennae.

We can predict some things from the cell geometry cross-section and its symmetries, illustrated in Fig.~\ref{fig:HOM-principle}.
If we ignore the coupling cell and the TE antenna placement on the DWG, we note that for the TE mode, the beam movement is perpendicular to the axis of the DWGs.
This implies that the distance between the DWG opening and the beam changes in the same way for both DWGs when the beam is moved in the plane the antennae are sensitive to, and we must have $c_L = c_R$.
For the TM mode this is not the case as the beam approaches one of antenna in a pair as it recedes from the other, and it is possible to have $c_L \ne c_R$.
Furthermore, the mirror-symmetry of the cell means that we should expect $c_L = - c_R$, where $x=0$ is the geometrical center of the cell.

This assumed symmetry is slightly broken by the placement of the TE antennas, which are displaced from the structure symmetry plane which is located in the center of the DWG to which they are attached, as can be understood from Fig.~\ref{fig:structure-antennas}.
This means that the signal center may not coincide with the geometrical center of the cell, and this effect can be different for the horizontal and vertical plane due to different antenna placement strategy.
Furthermore, as already mentioned in Section~\ref{sec:setup:RF}, the TE antennas sensitive to horizontal beam displacement (port~2 and~4) are located on opposite sides of their DWGs, giving a \SI{180}{\degree} relative phase shift, equivalent to $A_R = - A_L$.

\section{CST Simulation}
\label{sec:simu}

\begin{figure}
    \centering
    \includegraphics[width=0.99\linewidth]{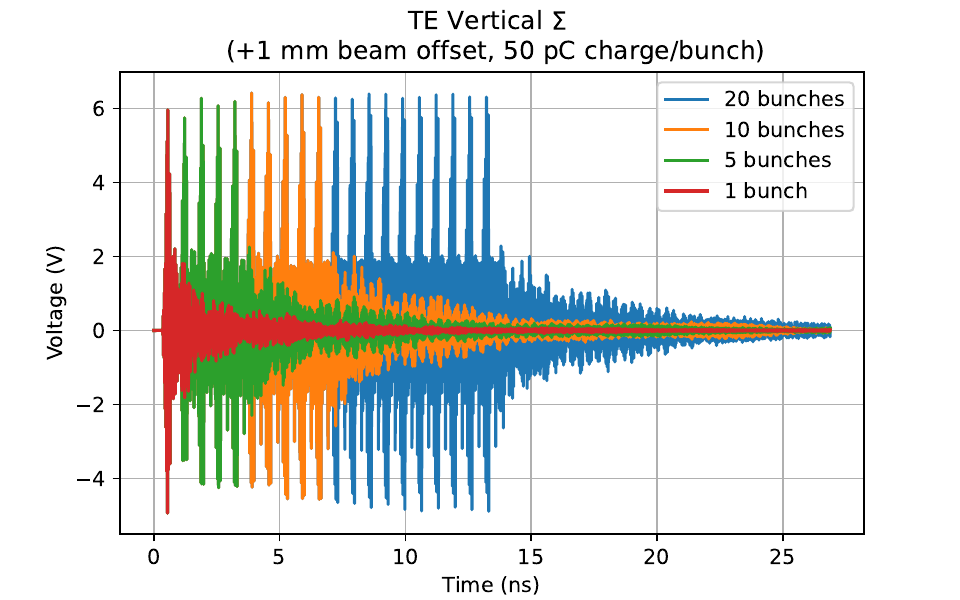}\\
    \includegraphics[width=0.99\linewidth]{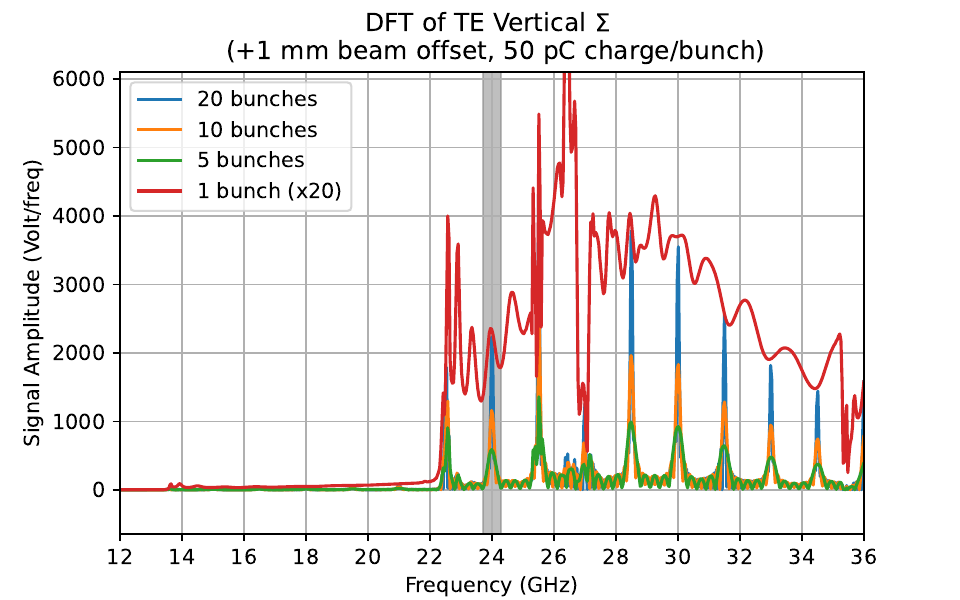}\\
    \includegraphics[width=0.99\linewidth]{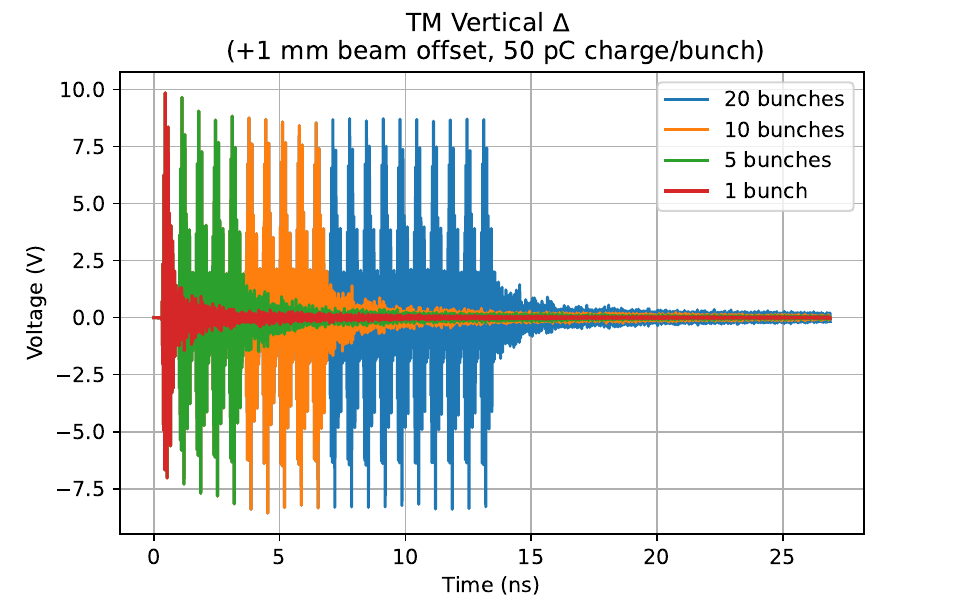}\\
    \includegraphics[width=0.99\linewidth]{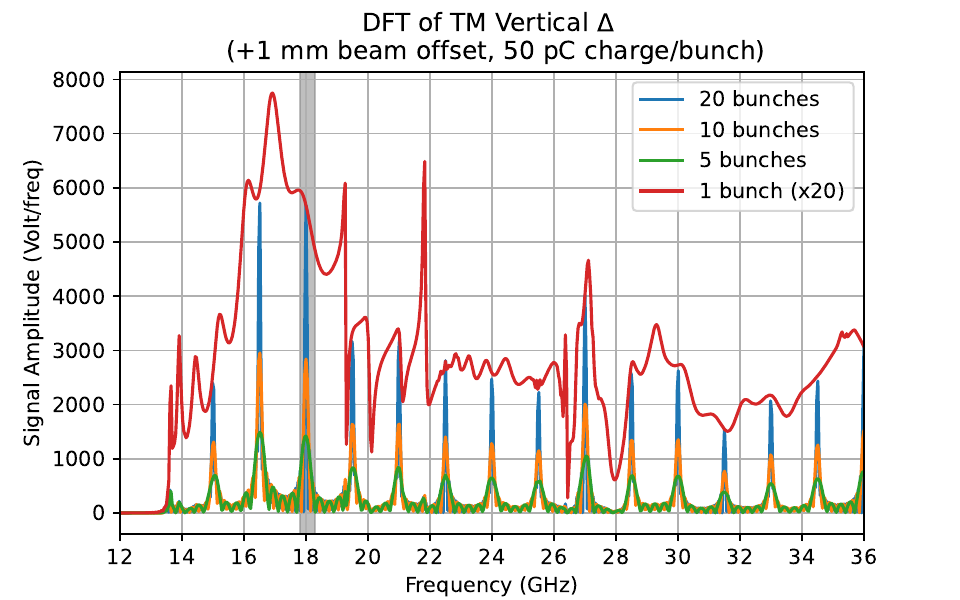}
    \caption{TE vertical sum and TM vertical difference signals in time and frequency domain for different number of bunches. Single-bunch DFT curve scaled $\times 20$ for readability.}
    \label{fig:simu:WFMsignals-multibunch}
\end{figure}

A simulation model was prepared in order to investigate the possible reasons for the discrepancy of the minimum signal positions between TE and TM WFM signals.
For this, an RF model of the TD26 structure~\cite{grudiev_geometry_2011} was imported to CST Particle Studio version~2020, as shown in Fig~\ref{fig:structure-antennas}.
In order to reduce the computing time by reduction of the computing domain, the extensions of the waveguides that carry the WFMs, and the waveguides for power coupling were both truncated to the length of the standard DWGs.
This necessitated moving the simulated antennae closer towards the center of the cell, placing the center of the antenna pins 75~mm from the center of the structure.
The antennas have a pin radius of 0.76~mm and an opening diameter of 1.75~mm; the pins penetrate 1~mm into the DWG for the TM and 0.8~mm for the TE antennas~\cite{solodko_rf_2012}.
Furthermore, the structure was shortened to just the input coupler, the DWG cell, and two regular cells.
This simplified geometry was found to be sufficient by a signal convergence study; a solver time step of 124~fs was used.

The simulations were performed using CST Wakefield Solver.
A Gaussian-shaped single bunch with a charge of 50~pC and an RMS length of 1~mm, corresponding to 3.3~ps, was used for all simulations.
The beam excited transverse wakefield signals were taken by the WFM antennas, giving a time-domain response for a single bunch.
Vertical beam offsets between -1 and +1~mm were used to the simulation model.

The processing of the simulated WFM signals followed the WFM data acquisition electronics scheme shown in Fig.~\ref{fig:wiring-schematics}.
After creation of $\Sigma$ and $\Delta$ signals in time domain, the results are converted to frequency domain using the Discrete Fourier Transform.
Bunch trains with 1.5~GHz spacing, as in CLEAR, are then created by using the time-shifting property of the DFT~\cite[pp.\ 373--375]{alan_v_oppenheim_signals_1996}
\begin{equation}
   \left[ x(t) \overset{DFT}{\leftrightarrow} X(\omega) \right] \to \left[ x(t+\Delta t) \overset{DFT}{\leftrightarrow} e^{-j\omega \Delta t} X (\omega) \right]
\end{equation}
on the single-bunch signals $x(t)$ and adding the results; this approach works even if the time shift $\Delta t$ is not an integer multiple of the sampling step.
The bunch harmonics are increasingly visible with more bunches, as seen in Fig.~\ref{fig:simu:WFMsignals-multibunch}.
In time domain, the signals are quickly decaying between the arrival of each bunch, due to the low Q-factors of the structure's well-damped HOMs.

\begin{figure}
     \centering
     \includegraphics[width=\linewidth]{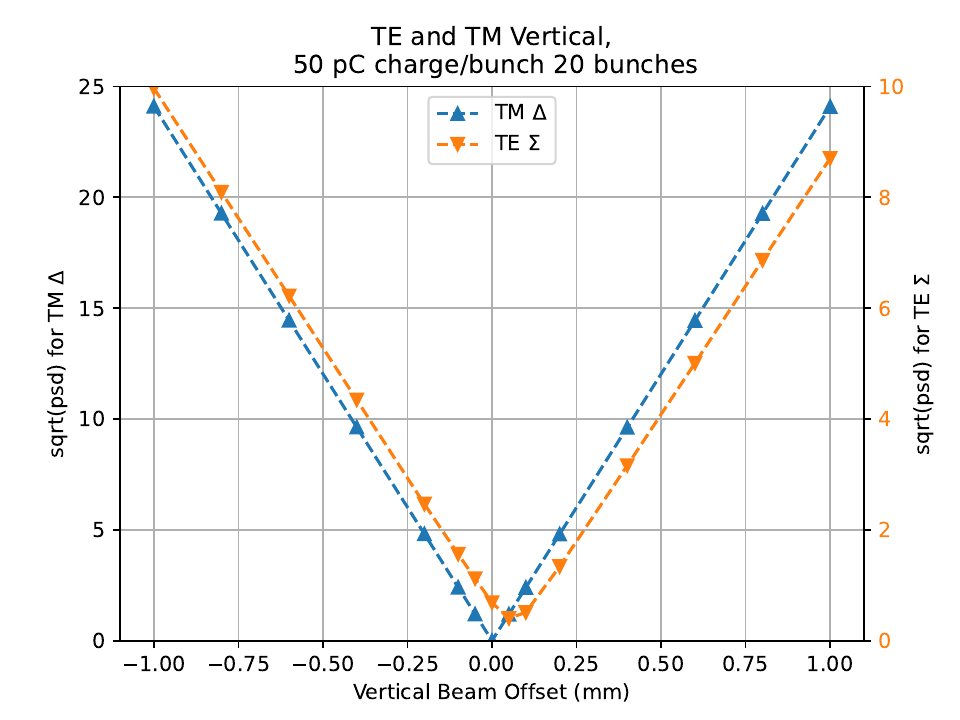}
     \caption{Simulated TE vertical $\Sigma$ and TM vertical $\Delta$ pulse integrals as function of the vertical beam position, using the standard antenna placement shown in Fig.~\ref{fig:structure-antennas}.}
     \label{fig:simu:V-TETM}
\end{figure}

Next, an ideal filter approach was applied to the TE and TM bunch train signals in the range of 23.7--24.3 GHz and 17.8--18.3 GHz, respectively, by setting the signals outside of the filter bandwidth to 0.
The power spectral densities (PSD) $|X(f)|^2 / N$ of the filtered signals are obtained by taking their power and normalizing with the number of DFT points $N$~\cite[Ch.~11]{ronald_n_bracewell_fourier_1999}.
The voltage is then found as the square root of this, averaged within the filter bandwidth, which up to a constant factor is equivalent to $\bar V$ from Eq.~\eqref{eq:signalIntegration}.
Both the TE and TM signals show a clear V-shaped response, as seen in Fig.~\ref{fig:simu:V-TETM}, similar to the experimental data in Fig.~\ref{fig:V-hybrid-standard}.

\subsection{Antenna position asymmetry}
\label{sec:simu:antenna}
From the signals shown in Fig.~\ref{fig:simu:V-TETM} and $x_0$ extracted from fitting Eq.~\eqref{eq:Vfit} to these, we get that the minimum signal strength for TE is found at $y=\SI{65}{\micro\meter}$, while the TM is centered at $y=0$.
This behavior is caused by the orientation of the TE antennas of port~1 and~3, shown in Fig.~\ref{fig:structure-antennas} and discussed in Sec.~\ref{sec:signalformation}, although it is possible that the effect is exaggerated by the shifting of the antennas closer to the cell in the simulation model than in the real structure.
In order to test this, a model with modified geometry was created in which the antenna for port~3 TE was moved to the top of the DWG instead of on the bottom, so that port~3 and~1 are located on opposite sides, similar to how the TE antennae for horizontal beam displacement are placed for port 2 and~4.
Note that this reverses the polarity of the voltage from the antenna, making the $\Sigma$ and $\Delta$ signals switch roles.

\begin{figure}
    \centering
    \includegraphics[width=\linewidth]{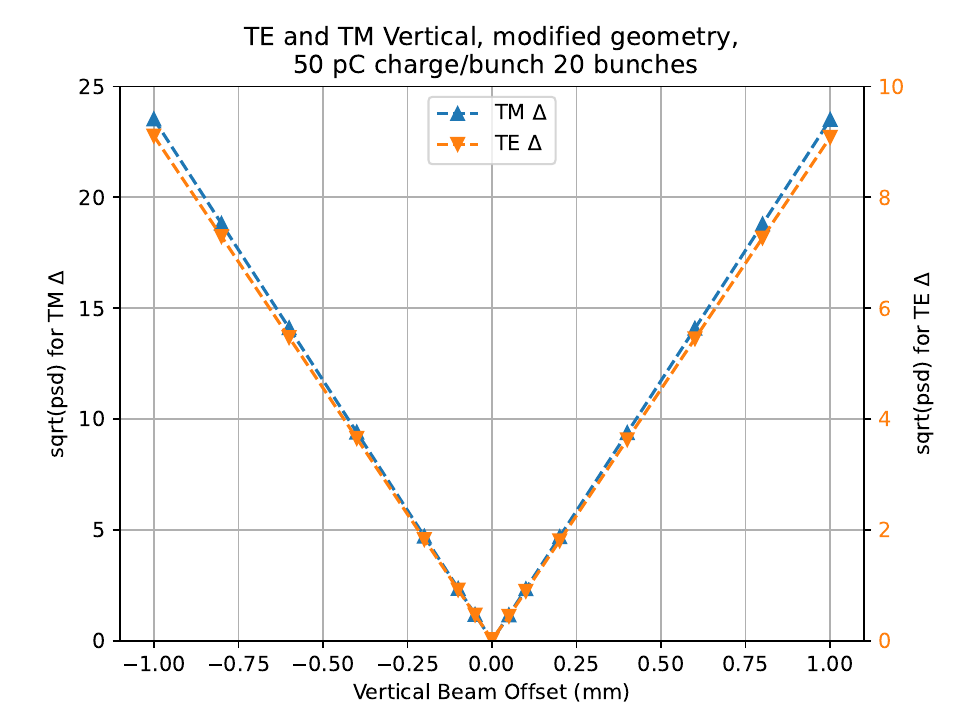}
    \caption{Simulated TE and TM vertical $\Delta$ pulse integrals, with the port~1 TE antenna moved to the other side of the DWG as discussed in Sec.~\ref{sec:simu:antenna}.}
    \label{fig:simu:WFMsignals-reversedAntenna}
 \end{figure}

The result of this simulation is shown in Fig~\ref{fig:simu:WFMsignals-reversedAntenna}.
The effect of this is that the electrical center as seen by the TE channel is also moved to $y=0$, and that the bottom of the V is sharpened.
Again, the effect of this might be exaggerated by the antennae locations in the simulation.

\subsection{Attenuation asymmetry}
\label{sec:simu:attenuation}

\begin{figure}
    \centering
    \includegraphics[width=\linewidth]{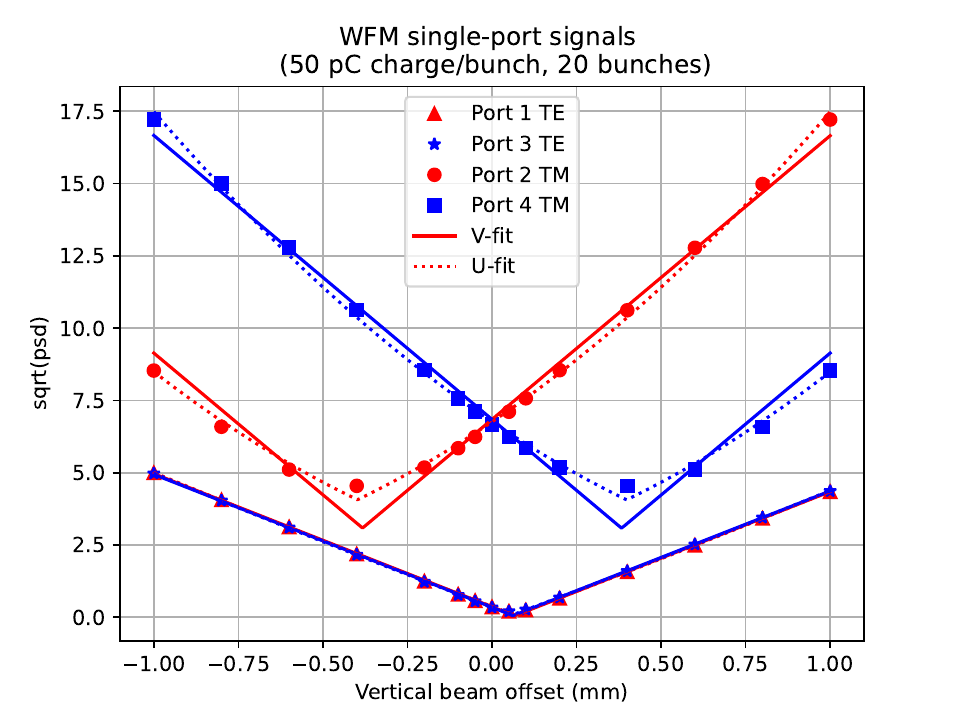}
    \caption{Simulated single-antenna signal amplitude as a response to vertical beam offset. Also indicated is V- and U-fits to the data.}
    \label{fig:simu:WFMsignals-attenuation}
\end{figure}

The predictions of the signal formation model described in Sec.~\ref{sec:signalformation} were tested using the simulated signals, which are generated from the standard simulation geometry and have a broad frequency spectrum.
The single-antenna voltages responses are shown in Fig.~\ref{fig:simu:WFMsignals-attenuation}, and $x_0$ from fits are shown in Tab.~\ref{tab:simu:WFMsignals-attenuation}.
Here V-fits means fits to Eq.~\eqref{eq:Vfit}, and U-fits are fits to
\begin{equation}
    V_\mathrm{fit} (x;A,B,x_0,V_0) = A \left|x-x_0 \right| + V_0 + B(x-x_0)^2\,,
    \label{eq:Ufit}
\end{equation}
which is also symmetrical around $x_0$ and is intended to also capture the soft ``bottom'' sometimes observed.
While these U-fits are fitting the shape of the simulation curves better, the resulting $x_0$ are almost identical.

\begin{table}
    \centering
    \begin{tabular}{c r l c | c}
        \hline
        Geom. & \multicolumn{2}{c}{Signal} & Fit type & $x_0$~[\si{\micro\meter}] \\
        \hline
        
        O & TE $\Sigma$,& no att.          & V &    65 \\
        O & TE $\Sigma$,& 50\% att. port 1 & V &    64 \\
        M & TE $\Delta$,& no att.          & V &     1 \\
        M & TE $\Delta$,& 50\% att. port 1 & V &   -19 \\
        \hline
        O & TM $\Delta$,& no att.          & V &     0 \\
        O & TM $\Delta$,& 50\% att. port 2 & V &   135 \\
        M & TM $\Delta$,& no att.          & V &     0 \\
        M & TM $\Delta$,& 50\% att. port 2 & V &   135 \\

        \hline
        \hline
        
        O & \multicolumn{2}{c}{Port 1 TE} & V &     67 \\
        O & \multicolumn{2}{c}{Port 1 TE} & U &     67 \\
        O & \multicolumn{2}{c}{Port 3 TE} & V &     62 \\
        O & \multicolumn{2}{c}{Port 3 TE} & U &     62 \\
        
        M & \multicolumn{2}{c}{Port 1 TE} & V &     62 \\
        M & \multicolumn{2}{c}{Port 1 TE} & U &     62 \\
        M & \multicolumn{2}{c}{Port 3 TE} & V &    -60 \\
        M & \multicolumn{2}{c}{Port 3 TE} & U &    -59 \\
        \hline
        O & \multicolumn{2}{c}{Port 2 TM} & V &   -383 \\
        O & \multicolumn{2}{c}{Port 2 TM} & U &   -398 \\
        O & \multicolumn{2}{c}{Port 4 TM} & V &    383 \\
        O & \multicolumn{2}{c}{Port 4 TM} & U &    398 \\
        
        M & \multicolumn{2}{c}{Port 2 TM} & V &   -384 \\
        M & \multicolumn{2}{c}{Port 2 TM} & U &   -399 \\
        M & \multicolumn{2}{c}{Port 4 TM} & V &    385 \\
        M & \multicolumn{2}{c}{Port 4 TM} & U &    410 \\

        \hline
    \end{tabular}
    \caption{Fitted centers $x_0$ of simulated pulse integrals as function of beam position, as shown in Figs.~\ref{fig:simu:V-TETM}, \ref{fig:simu:WFMsignals-reversedAntenna} and~\ref{fig:simu:WFMsignals-attenuation-combined} (above double horizontal line) and Fig.~\ref{fig:simu:WFMsignals-attenuation}  (below double horizontal line), all rounded to the nearest \si{\micro\meter}.
    Note that results from multiple subsections have been grouped together for convenient comparison.
    Results from signals with different TE antenna placements, as discussed in Sec.~\ref{sec:simu:antenna}, are indicated with original~(O) and modified~(M).
    Combined signals with attenuation asymmetry, discussed in Sec.~\ref{sec:simu:attenuation}, are also shown and compared to combined un-attenuated signals.
    The centers $x_0$ are estimated using fits with V-type (Eq.~\eqref{eq:Vfit}) and U-type functions (Eq.~\eqref{eq:Ufit}).}
    \label{tab:simu:WFMsignals-attenuation}
\end{table}

It is here clearly visible that for TM, the two antennas used for the same mode do not have the same minimum.
As discussed earlier, this is a sufficient condition for the combined signal to be off-center if there is asymmetric attenuation.
For TE, the two antennas do have the same minimum, however it is shifted to the side, explaining the results in Sec.~\ref{sec:simu:antenna}.
The table also shows the fitted centers for the modified geometry, in which one of the TE antennas see a change of sign.

\begin{figure}
    \centering
    \includegraphics[width=\linewidth]{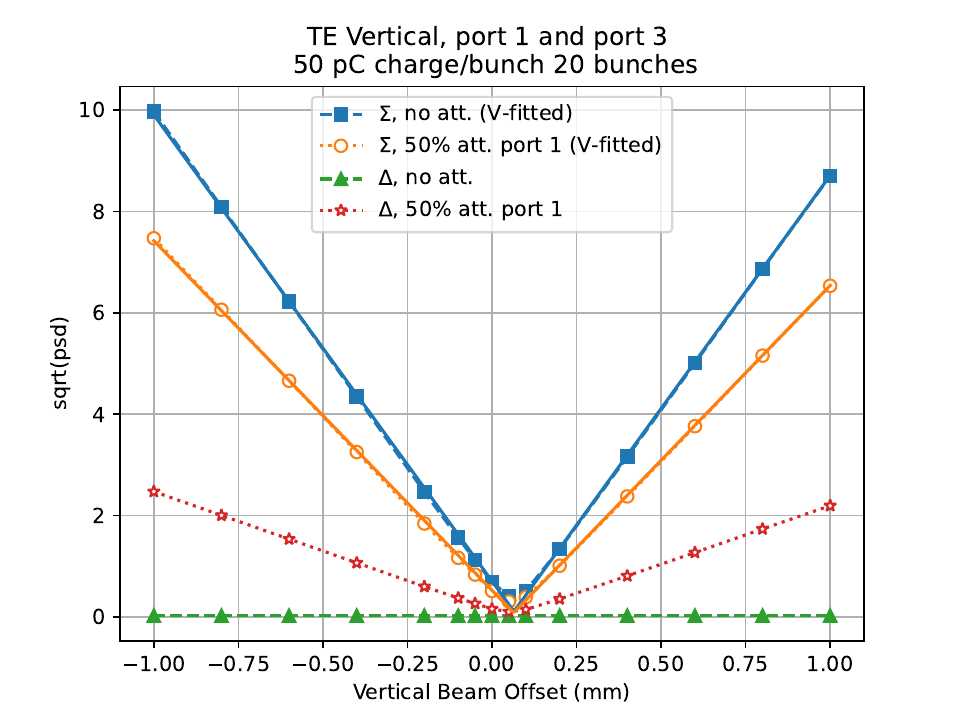}\\
    \includegraphics[width=\linewidth]{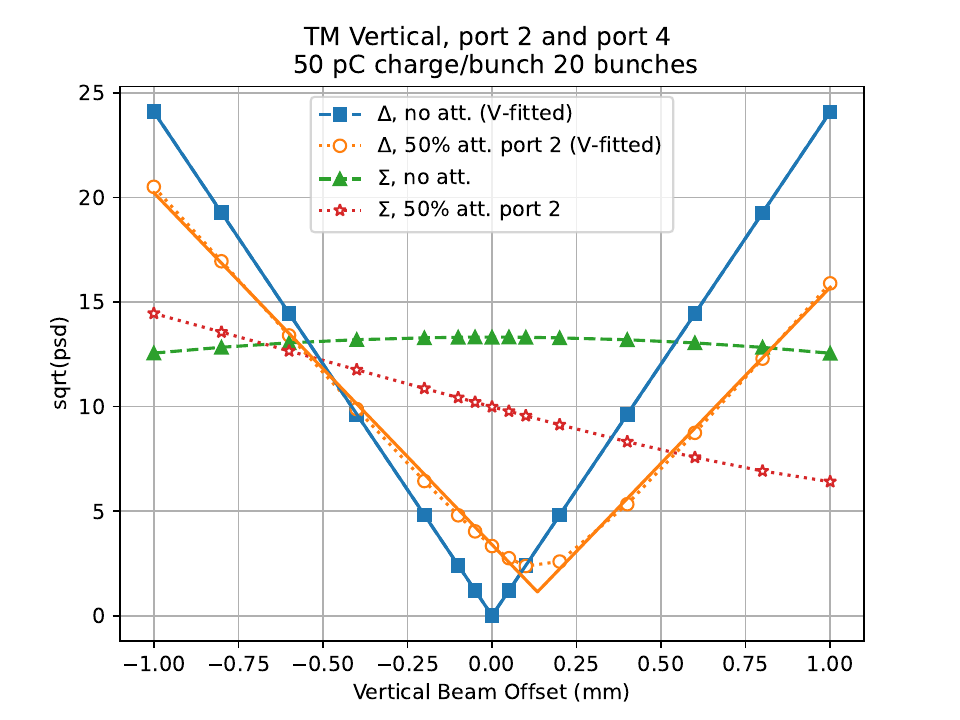}\\
    \caption{Simulated effect of attenuation asymmetry on the combined $\Sigma$ and $\Delta$ signals with vertical beam offsets, for TE (top) and TM (bottom), including V-fits.}
    \label{fig:simu:WFMsignals-attenuation-combined}
\end{figure}

Combining the signals as shown in Fig.~\ref{fig:simu:WFMsignals-attenuation-combined} demonstrates the effect of asymmetrical attenuation; here 50\% attenuation is applied to port 1 (TE vertical signals) or port 2 (TM vertical signals).
The result is a shift of the center towards the attenuated antenna, and also a softening of the center.
Furthermore, some power is shifted into the non-signal arm of the hybrid, i.e. into the $\Delta$ arm of TE signal in the original geometry, and the $\Sigma$ arm of the TM signal.

As can be seen from Tab.~\ref{tab:simu:WFMsignals-attenuation}, the effect is different for the TE signal in the modified geometry, where both antennas have the same center, which is different from the structure center.
This means that the center of the combined signal does not shift when the attenuation becomes asymmetrical.

\subsection{Bunch train distortion}
\label{sec:simu:distortion}

\begin{figure}
    \centering
    \includegraphics[width=\linewidth]{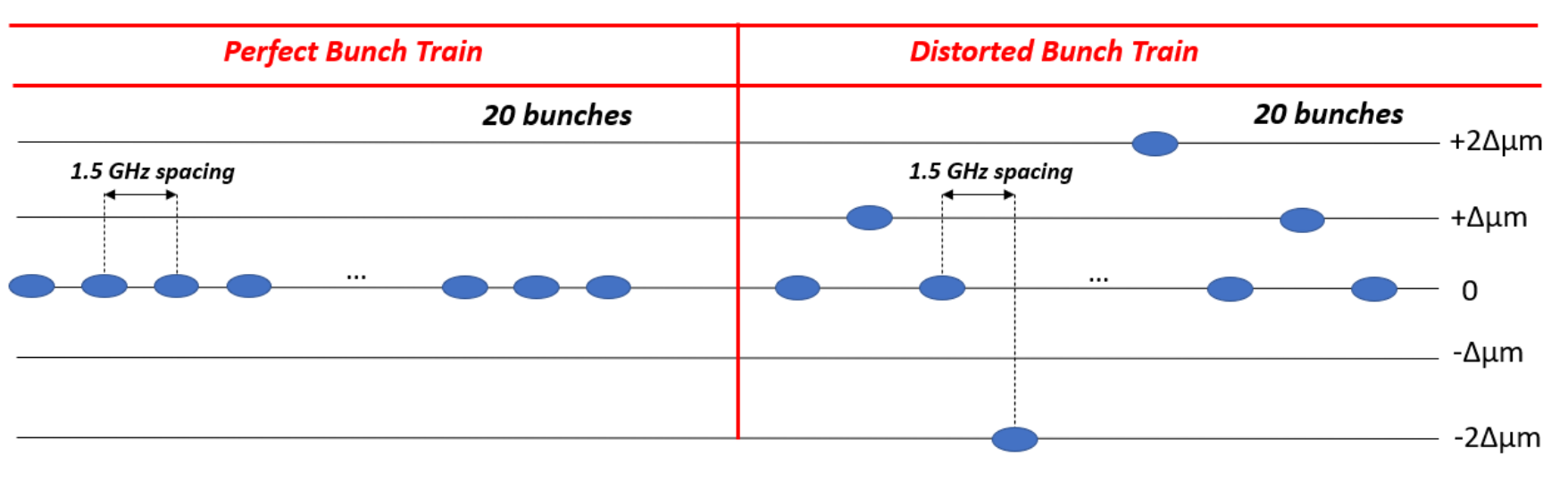}
    \caption{Perfect vs.\ distorted bunch trains, where $\Delta$ is the vertical spacing between the wakefield simulation points from which the bunch positions are drawn.}
    \label{fig:simu:bunchTrain-distorted}
\end{figure}

Another considered source of shifts in the electrical center of each mode, is random offsets of the bunch position from the average of the train, as illustrated in Fig.~\ref{fig:simu:bunchTrain-distorted}.
In order to simulate this, bunch trains were created by combining signals from bunches with different center offsets.
This was done by drawing from a pool of pre-simulated wake fields with different offsets, computed at every $\Delta=\SI{50}{\micro\meter}$ from $y=\SI{-500}{\micro\meter}$ to $y=+\SI{500}{\micro\meter}$.
The bunch offsets were drawn from a Gaussian distribution with a standard deviation $\sigma_b$, where each offset was rounded to the nearest simulation point and the distribution was truncated at $\pm\SI{300}{\micro\meter}$ to stay within the generated range when sweeping.
Each generated bunch train was then swept across the structure from $y=\SI{-200}{\micro\meter}$ to \SI{+200}{\micro\meter} without re-shuffling the bunches within one sweep.
It is expected that the bunch offsets are caused by reproducible effects resulting from dispersion and long-range wake-fields in the upstream injector linac.
For each sweep of the beam, $\bar V(y)$ was fitted to Eq.~\eqref{eq:Vfit}, and the fitted center position $x_0$ was compared to the true centroid offset position $y_c$ of the generated beam.
It was noted that when $\sigma_b>0$, the bottom of the TM channel became ``rounder'' in the same manner as seen in Fig.~\ref{fig:simu:WFMsignals-attenuation-combined}~(bottom).
Note that $y_c$ is generally nonzero since it is the average of 20 sampled bunch offsets, so that it is itself distributed with $\sigma=\sigma_b/\sqrt{20}$.

\begin{figure}
    \centering
    \includegraphics[width=\linewidth]{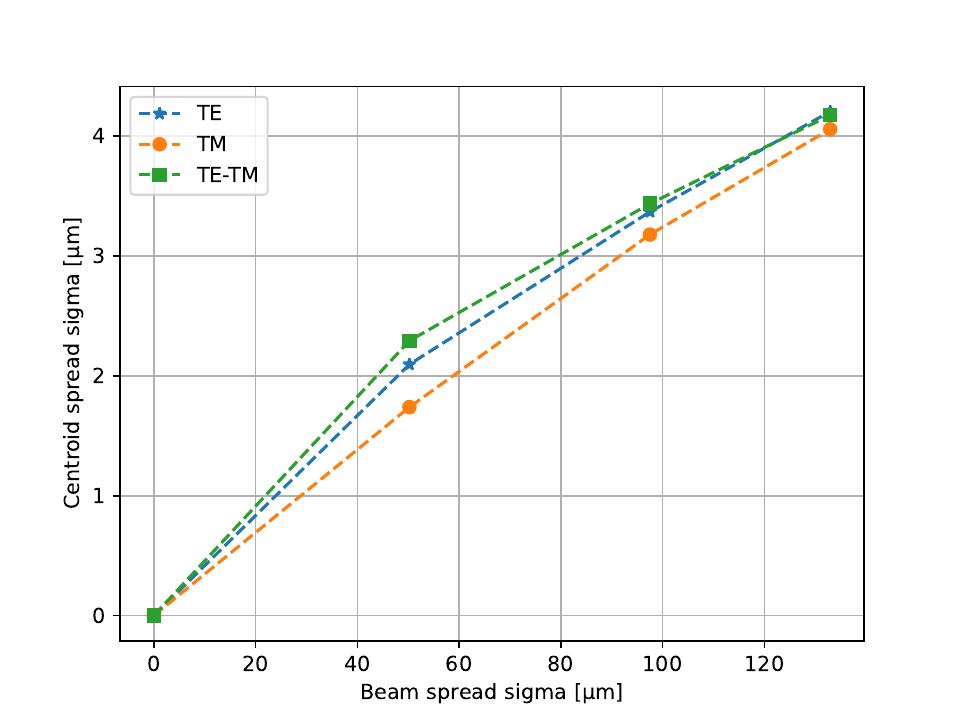}
    \caption{Error in the estimated structure center $x_0$ relative to beam centroid offset $y_c$ due to distortion of simulated bunch trains, as discussed in Sec.~\ref{sec:simu:distortion}.}
    \label{fig:simu:distortionError}
\end{figure}

In order to statistically study the effect of the bunch train distortion quantified as $\sigma_b$, 1000 different bunch trains were generated and swept across the structure for each $\sigma_b$.
The variation of $y_c-y_0$ was quantified as the standard deviation of this quantity, and plotted as a function of $\sigma_b$ in Fig.~\ref{fig:simu:distortionError}.
Also shown is the variation in the fitted $x_0$ between the TE- and TM-channels.
Note that the $\sigma_b$ in this figure is the sample standard deviation of the spread of the generated bunches, including the effect of truncation.
As seen in the figure, there is an approximately linear growth in the centroid discrepancy, at about \SI{3}{\micro\meter} per \SI{100}{\micro\meter}.

Additionally, simulations where the bunch charge was varied along each train were also conducted.
Here, the charge of each bunch relative to the average charge was varied with a Gaussian distribution with a standard deviation of 10\%, with the total charge kept constant between generated trains, and negative charges were disallowed.
This did not produce any clear effects on top of what is shown in Fig.~\ref{fig:simu:distortionError}.

\section{Experimental comparison of TM and TE channels}
Several experiments have been done in order to explore hypotheses about the underlying cause of the difference in the center position as measured by the TM and TE channels.
These are described in the subsections below.

\subsection{Attenuation asymmetry}
\label{sec:center-comparison:hybrid-modify}

\begin{figure*}
    \centering
    \begin{subfigure}{0.48\textwidth}
        \includegraphics[width=\textwidth]{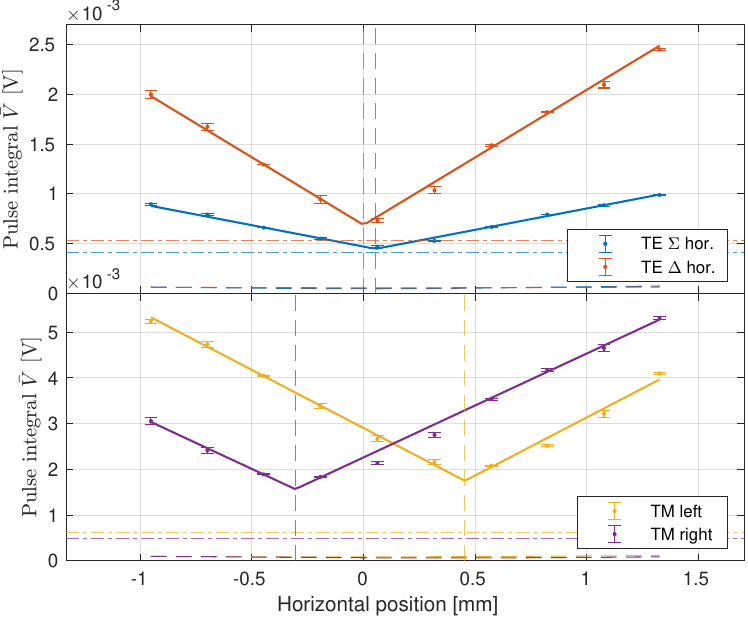}
        \caption{Horizontal scan, TM hybrids bypassed.}
    \end{subfigure}
    \begin{subfigure}{0.48\textwidth}
        \includegraphics[width=\textwidth]{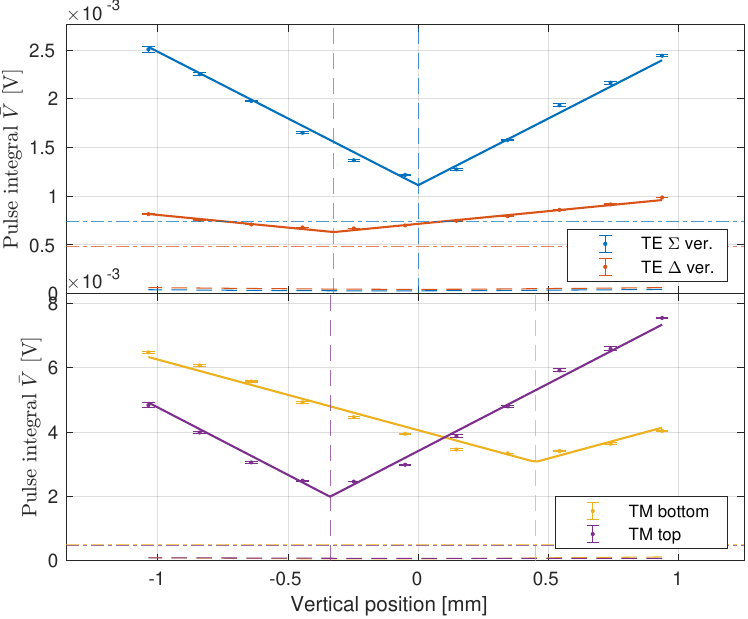}
        \caption{Vertical scan, TM hybrids bypassed.}
    \end{subfigure}
    \begin{subfigure}{0.48\textwidth}
        \includegraphics[width=\textwidth]{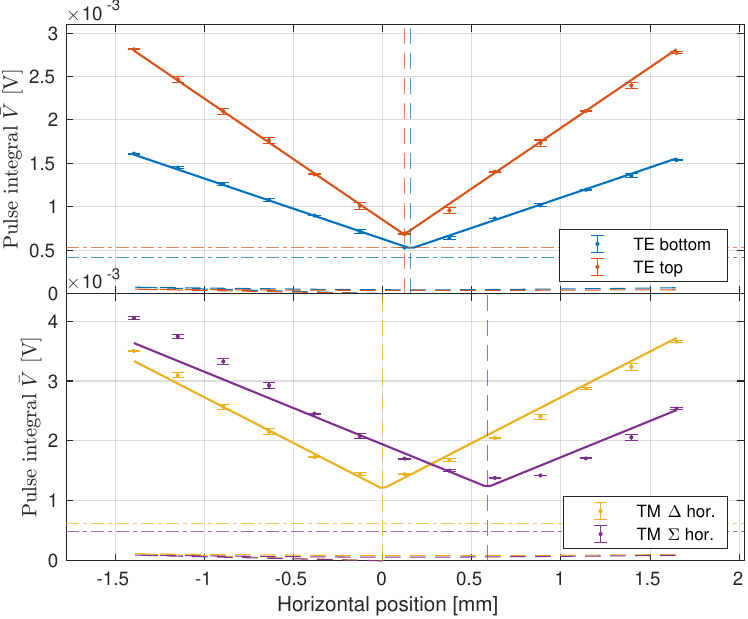}
        \caption{Horizontal scan, TE hybrids bypassed.}
    \end{subfigure}
    \begin{subfigure}{0.48\textwidth}
        \includegraphics[width=\textwidth]{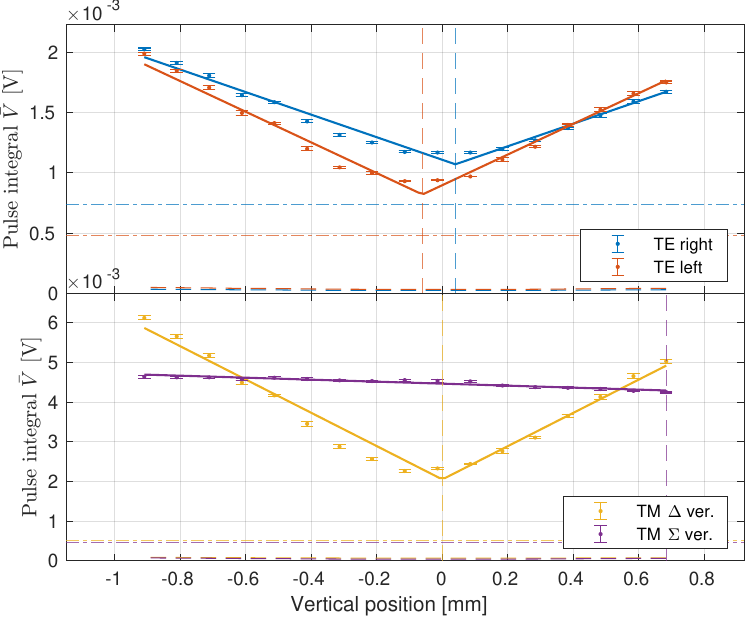}
        \caption{Vertical scan, TE hybrids bypassed.}
    \end{subfigure}
    \caption{Attenuation asymmetry experimental test described in Section~\ref{sec:center-comparison:hybrid-modify}, plotted and fitted in the same way as in Fig.~\ref{fig:V-hybrid-standard}.
    Subfigures (a,b) shows the effect of bypassing the hybrids for TM signals, while (c,d) shows the effect for bypassing the hybrids for TE signals.
    The plots are centered so that $x_0=0$ for the ``reference'' un-bypassed channel for each scan. Increasing vertical position means that the beam is moving upwards, increasing horizontal position means that the beam is moving to the right.}
    \label{fig:V-hybrid-modify}
\end{figure*}
\begin{table*}
    \centering
    \caption{Fitted center positions $x_0$ in \si{\micro\meter} using Eq.~\eqref{eq:Vfit}, from the experimental data shown in Fig.~\ref{fig:V-hybrid-modify}.}
    \label{tab:V-hybrid-modify}
    \begin{tabular}{c c c | c c | c c}
        \hline
        \hline
        Plane & Mod.\ ch & Subplot & Mod.\ Left/Bottom & Mod.\ Right/Top & Ref.\ $\Sigma$ & Ref.\ $\Delta$ \\
        \hline
        H & TM & (a) &  $454\pm4$ & $-306\pm3$ &  $54\pm3$ &       $0\pm3$ \\ 
        V & TM & (b) &  $450\pm3$ & $-339\pm2$ &   $0\pm2$ & $-0.324 \pm2$ \\ 
        H & TE & (c) &  $158\pm2$ &  $121\pm2$ & $588\pm3$ &       $0\pm2$ \\ 
        V & TE & (d) &  $-61\pm2$ &   $40\pm3$ & $683\pm3$ &       $0\pm2$ \\ 
        \hline
        \hline
    \end{tabular}
\end{table*}

As discussed in Sec.~\ref{sec:signalformation} and explored through simulation in Sec.~\ref{sec:simu:attenuation}, if the signal minimum for the antennas in a pair do not coincide, the combined center is sensitive to imbalances of the signal attenuation before the hybrid.
In order to test this experimentally, the hybrids were removed on TE- or TM-channels, and the signals directly from the antennas were connected to the band-pass filters.
In each case, the signals from the other channels were connected normally and used as a reference.
The structure was then shifted across the beam.
The results from this is shown in Fig.~\ref{fig:V-hybrid-modify}, and in Tab.~\ref{tab:V-hybrid-modify}.

Note that between the runs with TE bypassed and those with TM bypassed, we needed to stop the accelerator for an access to modify the wiring; because of this the beam conditions and especially the absolute beam positions are not identical for the two data sets.
Both scans were run with 32 bunches with an average charge of 115~pC and bunch length of 3.5~ps.
Since the absolute beam position was unknown, the horizontal axis on the plots are relative to the primary channel of the mode where hybrid was not bypassed.

As we see from the data in Tab.~\ref{tab:V-hybrid-modify} and Fig.~\ref{fig:V-hybrid-modify}, the center positions of the single-antenna TM-modes are split and on different sides of the TE modes, with an average half-distance of 380 and \SI{396}{\micro\meter}.
On the contrary for TE modes, when disconnected from the hybrid they stay close together, giving a center position which is comparable to the position from the reference TM mode.
This supports the hypothesis from Section~\ref{sec:signalformation}, that due to the symmetry of the cell, the TE signals have less of a center shift than the TM signals.
Furthermore, we see that the minimum for the TM signals are displaced away from the antennae, as also expected.

\subsection{Beam charge and quality}
\label{sec:center-comparison:charge-scan}

\begin{figure}
    \centering
    \includegraphics[width=\linewidth]{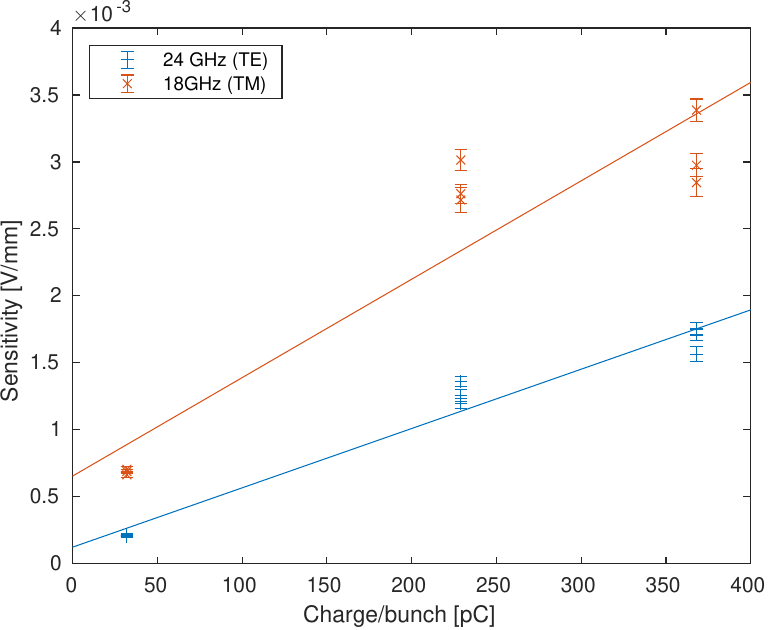}
    \caption{Strength of WFM response as a function of bunch charge, quantified by $A$ from in Equation~\eqref{eq:Vfit}, together with linear fits.}
    \label{fig:chargeScan-response}
\end{figure}
\begin{figure*}
    \centering
    \includegraphics[width=0.32\linewidth]{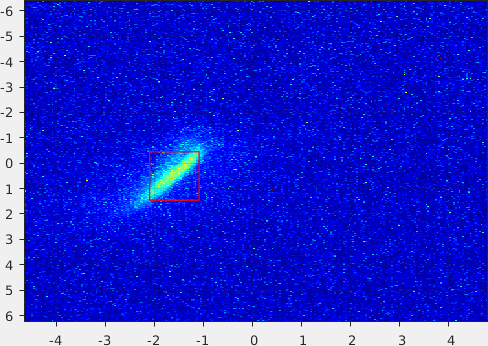}
    \hfill
    \includegraphics[width=0.32\linewidth]{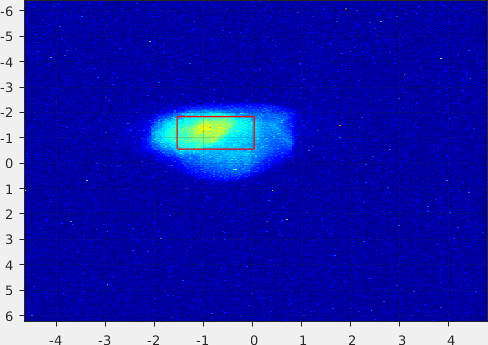}
    \hfill
    \includegraphics[width=0.32\linewidth]{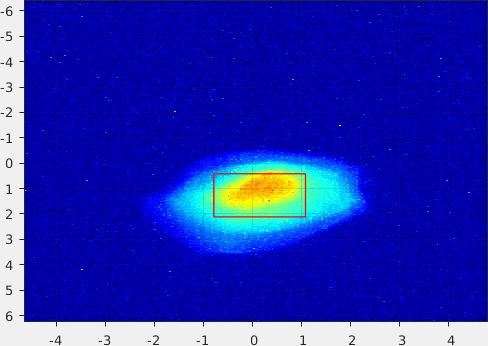}
    \caption{Beam profile images at the first profile monitor at CLEAR (BTV390) with 20 bunches and (from left to right) 23, 229, and 268~pC average bunch charge. Scales are in mm. The red boxes indicate the region used for bunch profile fits in the software used to extract the images. The beam is magnified on this screen by the magnetic optics compared to when passing the WFM system in the structure.}
    \label{fig:chargeScan-profiles}
\end{figure*}
\begin{figure}
    \centering
    \includegraphics[width=\linewidth]{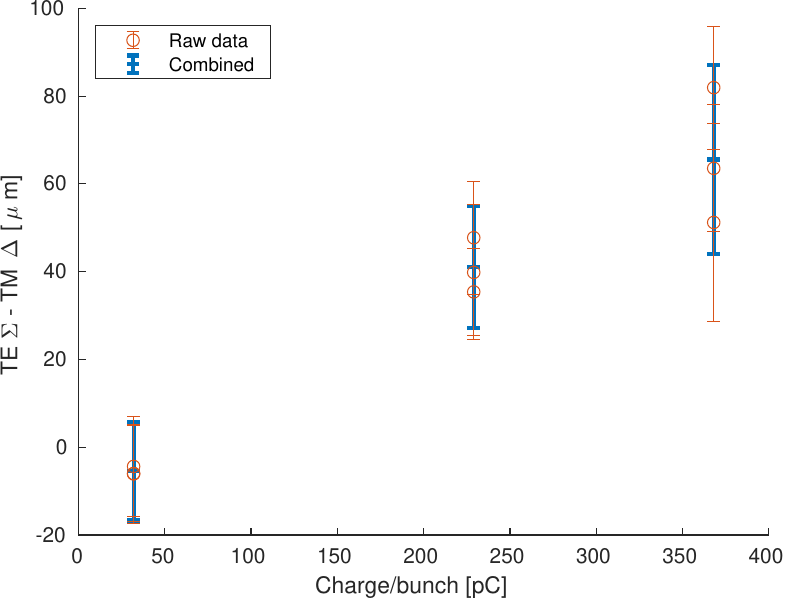}
    \caption{Difference in TM and TE vertical center position as a function of bunch charge, using the same scans as in Fig.~\ref{fig:chargeScan-response}.}
    \label{fig:chargeScan-center}
\end{figure}

In order to verify the operation at different bunch charges, the beam charge was varied while the number of bunches and bunch length was kept constant at 20~bunches and 3~ps bunch length, with a beam momentum of 211~MeV/c.
Three immediately consecutive scans were performed at each charge for error estimation.
As shown in Fig.~\ref{fig:chargeScan-response}, the response increases with charge, however the increase is not exactly linear.

When the charge per bunch is increased, an enlargement of the beam size on the screens upstream of the structure is clearly visible, as shown in Fig.~\ref{fig:chargeScan-profiles}.
When the upstream YAG screen is viewed with strong filters the bunches can be seen as separated points.
This is expected, since higher bunch charges induce larger wake fields in the upstream C-band linac, which in turn kicks the later bunches in the beam.
The bunch charge can therefore be used as a proxy to the beam quality, with higher charges corresponding to increased bunch train distortion.

From the data in Fig.~\ref{fig:chargeScan-center}, we see that the difference in center position measured by the TE- and TM-pickups also increase with bunch charge.
Here, the figure shows both the difference in $x_0$ for TE and TM, as well as a combined estimate from all scans at a given charge using a mixture distribution from all three points at each charge.
Since the signal production should be linear with bunch charge if the charge distribution is the same, this indicates that the two channels respond differently when the beam quality deteriorates.

\subsection{Beam angle}
\label{sec:center-comparison:beam-angle}

\begin{figure}
    \centering
    \includegraphics[width=\linewidth]{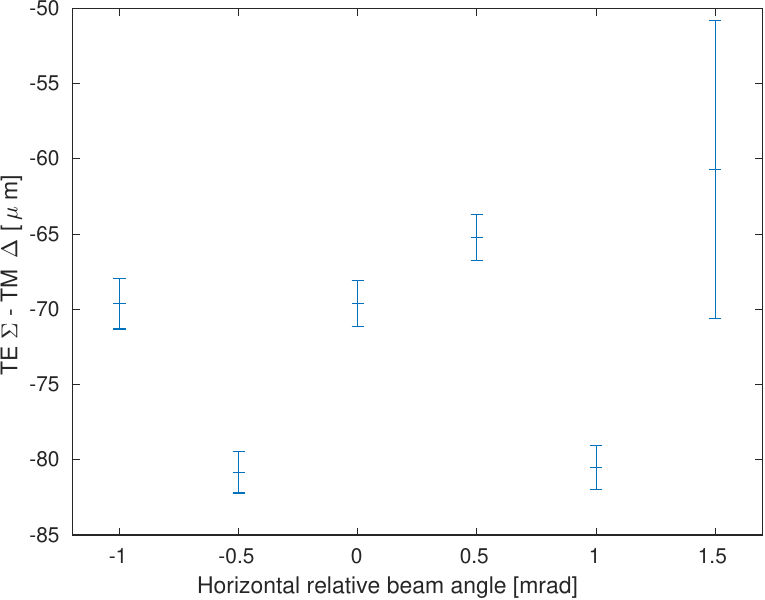}
    \caption{Difference in horizontal center position $x_0$ measured by TE and TM modes, as a function of beam angle.}
    \label{fig:angleScan}
\end{figure}

One hypothesis was that the observed variability between the relative center of the TE- and TM-signals was caused by a beam angle.
Due to the lack of an accurate BPM system at CLEAR, the beam angle is not exactly reproducible between shifts or when manipulating other parameters such as the beam charge or bunch length.
This makes a variation in beam angle a possible explanation of the variable discrepancy between the two channels.
The effect of this was evaluated through the use of the bump scan method~\cite{gilardi_measurements_2021}, where the beam angle can be controlled by using slightly different kick strength on the two corrector magnets.
The beam was scanned over the structure aperture using the same magnets also used to create the beam angle.
The beam angle variation was set up so that at for a given beam position setting, the beam position was fixed and independent of the angle setting on a screen just upstream of the structure.
This meant that the absolute position of the beam in the structure varied slightly between scans of different angle.
Again, the observable of interest is the split in $x_0$ between the TM and TE channels.

For each scan, we measured the distance between the minimum the V in the TM and the TE channel.
This is plotted in Fig.~\ref{fig:angleScan}, showing no clear correlation, indicating that the beam angle does not significantly contribute to the variation of the discrepancy.
The variation is also small, on the order of \SI{10}{\micro\meter}.

For this scan, a train of 50~bunches and a total charge of 11.7~nC (corresponding to a bunch charge of 234~pC) was used, a bunch length of 7~ps, and an average particle momentum of 195~MeV/c.

\section{Discussion}

As stated in the introduction and seen in Fig.~\ref{fig:V-hybrid-standard}, there is often a discrepancy in the position $x_0$ of the minimum signal as observed in the TE- and TM-bands in this structure.
We have shown that this can be caused by both a difference in the position $c_{L/R}$ of the minimum signal as measured by each antenna in each pair.
This especially affects the TM-band, as explored in Sections~\ref{sec:signalformation}, \ref{sec:simu:attenuation}, and~\ref{sec:center-comparison:hybrid-modify}.
However, as seen in Sec.~\ref{sec:center-comparison:charge-scan} the difference is not constant, and from simulation it is not explained by bunch train distortion as seen in Sec.~\ref{sec:simu:distortion}.
Finally, as seen in Section~\ref{sec:center-comparison:beam-angle} there is no correlation between the beam angle and the observed difference between the channels.

This presents a challenge for using the WFM system to align the structure relatively to the beam trajectory so that the wake field kick is minimized.
Especially, it requires that either the WFM attenuation and phase of the two sides on every structure is carefully balanced, or that the system on every structure is calibrated relatively to a reference; both of these alternatives is time-consuming and depends on the attenuations and phase shifts remaining stable.
Even then, it appears that the center of the combined signals from the hybrids are not stable when changing beam conditions.

The main difference between this WFM system and a typical resonant cavity BPM system, is that in this system we do not have one principle mode that is excited and measured.
The radiation that reaches the antennas and pass the filters are therefore likely a mixture of radiation from multiple modes with different field patterns, and the diffraction of the field around the electron bunches into the damping wave guides.
This complicated signal formation process does not appear to lean itself well to a hybrid-based RF front-end.

The most direct comparison we can do between the simulations and the experimental data is for the single-port data, without hybrids which are affected by attenuation asymmetries.
For this, we look at the fitted centers in Tabs.~\ref{tab:simu:WFMsignals-attenuation} (original geometry) and~\ref{tab:V-hybrid-modify}, which is derived from the data presented in Figs.~\ref{fig:simu:WFMsignals-attenuation} and~\ref{fig:V-hybrid-modify}.
This shows that the TE signals show almost the same center for both antennas, with only \SI{5}{\micro\meter} of separation in simulation and \SI{21}{\micro\meter} in the experimental data from the vertical scan.
For the TM signals we have \SI{776}{\micro\meter} in simulation and \SI{789}{\micro\meter} in the experimental data.
This means that the difference between simulation and experiment is on the order of just \SI{15}{\micro\meter}.
Along with the qualitative agreement of the curve shapes in Figs.~\ref{fig:simu:WFMsignals-attenuation} and~\ref{fig:V-hybrid-modify}, as well as Figs.~\ref{fig:V-hybrid-standard}, \ref{fig:V-simplemodel}, and~\ref{fig:simu:WFMsignals-attenuation-combined}, this strongly indicates that modeling $c_L \ne c_R$ for the TM signal is correct, and that it can cause a shift in the center as observed in the TE- and TM signals.

The difference in $c_L$ and $c_R$ could be used in a beneficial manner:
If no hybrid is connected and two acquisition channels are used for each plane on each structure, it allows the determination of the sign of the beam displacement in addition of the magnitude, something that is not possible with the originally foreseen system using a hybrid and a single acquisition channel.
Furthermore, the minima seen by each antenna would then be determined by the interior geometry of the cell and not by the signal attenuation in the cabling of the structure, which should greatly ease calibration.
Finally, the precision of the device when the beam is near center, which is where it needs to be most precise, should also be improved in this mode of operation, since the change in signal strength with respect to beam displacement will not be zero.
The stability of this mode of operation has however not been demonstrated.

In addition to the center-shift caused by attenuation, a small systematic shift of the TE center even after a perfectly balanced hybrid is also evident in the simulation.
This was found to be caused by the positioning of the antennae inside the damping waveguides; it should be noted that this effect is expected to be of a different magnitude in reality than in simulation, since in the simulation the DWGs have been shortened.

In general the TE-band was much less sensitive to the effect of attenuation imbalance due to the symmetry of the positioning of the damping waveguides that capture the beam position signal, relatively to the plane of beam motion.
This indicates that if a single channel and a hybrid (or even a single antenna) is to be used, it is likely preferable to use the TE mode.

An important challenge for the WFM system investigated here is that discrepancy between the TE and TM signals is not stable, as seen when changing the beam charge in Sec.~\ref{sec:center-comparison:charge-scan} but keeping the train- and bunch lengths constant.
This was also explored in simulation in Sec.~\ref{sec:simu:distortion}, where an effect of approximately \SI{3}{\micro\meter} per \SI{100}{\micro\meter} was observed.
This is difficult to compare directly to the experimental data since no profile monitor can be installed exactly at the location of the WFM, however extrapolation indicates that the effect would be at the experimentally observed level with a $\sigma_b\approx\SI{1}{\milli\meter}$.
This is large, but not implausible for the experimentally used trains with high bunch charges.

Finally, based on the results from changing the beam angle as discussed in Sec.~\ref{sec:center-comparison:beam-angle}, it is unlikely that there is any contribution from the beam angle changing with beam conditions.

Given the results presented for the current CLIC wake field monitor design, it appears prudent to explore structures with modified geometries at the location of the integrated beam position monitor.
This has been done in the past, using a superstructure consisting of two CLIC-type accelerating structures with WFMs installed in the middle cells of both structures~\cite{quirante_califes_2014,farabolini_recent_2014}.
In that setup, a precision better than \SI{5}{\micro\meter} was observed in the TE mode.

The TD24 structure has the WFM pickups located in the middle of the structure, while TD26 structure also used in this paper have the pickups mounted in the first cell of the structure.
Furthermore, while each cell of the TD24 structure was machined with damping wave guides, unlike for the TD26 structure tested in this paper, they were not terminated in RF loads.
Instead, simple holes were drilled through the end of the normal DWGs for vacuum purposes, venting the structure into the enclosing vacuum tank.
This created partial RF shorts at the ends of the DWGs of the TD24, which is unlikely to be perfectly balanced between pairs of waveguides.
As a result, the center of the higher order modes of this structure are likely to vary throughout it, and be determined by the geometry of the DWGs as well as the accelerating cell and iris.
Since the experimental TD24 structure was not well damped, it would also mean that wake-excited field could propagate longer than in the TD26 structure.
This would complicate interpretation of the results from the earlier tests of the TD24 structure, especially if interested in the accuracy of the position of the electromagnetic center of the kicks relative to the center of the measured wake field.

It was difficult to determine the resolution of the WFM system as presented in this paper, due to the lack of other high resolution beam position monitors nearby.
This means that any position estimate would also include contributions from beam position jitter, and that it can only be given for a scan, not for a single bunch.
We can however compare the measured $x_0$ between the TE- and TM channel in a scan, as in e.g.\ Figs.~\ref{fig:chargeScan-center} and~\ref{fig:angleScan}.
These show that the difference between the channels vary with approximately \SI{10}{\micro\meter} between scans using the same beam configuration.
While this is not a direct measurement of the resolution, it indicates that the underlying resolution might be comparable to what was observed in the previous structure.
Furthermore, the statistical error estimates for $x_0$ from single scans are on the order of 2-3~\si{\micro\meter}, as seen from Tabs.~\ref{tab:V-hybrid-standard} and~\ref{tab:V-hybrid-modify}.

\section{Conclusions}

We have observed a difference between the electrical center of  the TD26 CLIC accelerating structure as measured with the TE- and TM channel of the CLEAR WFM system to beam position.
Furthermore, this difference is seen to be variable with beam conditions.

Using a combination of beam measurements, theory, and computer simulation we have discovered three mechanisms that can create such a discrepancy.
The main mechanism is that the system is strongly sensitive to the balance of attenuation of the two antennas in each pair.
Furthermore, the slightly broken symmetry of the placement of the TE antennas create a small shift in the minima of the vertical TE channel.
Finally, we see a growth in the difference between the channels with increasing bunch train distortion $\sigma_b$, as well as a ``softening'' of the minimum of the TM signal.

This presents a challenge for the CLIC WFM system as currently designed, especially in the TM-band.
It may however be possible to overcome by using two acquisition channels measuring the power per plane and structure, one per pickup antenna, instead of a hybrid and a single acquisition channel.
Such a system would also avoid the loss of position sensitivity around the electrical center, which is otherwise inherent when trying to determine the structure center from the combined power level.
It would also be able to determine the sign of the beam displacement, which is not possible in the hybrid-based RF front-end design.
Another possible solution is a modification of the cell geometry at the location of the WFM.

For the TE channel, the geometric symmetry of the system is unbroken by the beam approaching one antenna more than the other, resulting in an almost centered voltage curve. 
This indicates that this channel may be preferable to use, and can even work with a single antenna and no hybrid.
Judging by the observed single-bunch frequency spectrum at the antenna, using a filter with a higher center frequency than the one tested here should give a stronger signal.

Finally, the results are in itself interesting in that they point out important issues with the type of beam position monitoring systems as are used here, where unlike with a typical resonant cavity BPM there is no strong mode that is being excited.
This could inform both future wake field monitor designs, and other beam position monitoring systems based on diffraction of the beam field through apertures in the beam pipe.

\begin{acknowledgments}
We thank CERN for providing the beam time at the CLEAR user facility.
Especially, we thank the alignment and survey team for helping us work with the original structure movement system and to verify its position relative to the rest of the linac, Samantha Pittman for helping us with the RF acquisition cabling, and Pierre Korysko for his work on the new remote position readout system.
This work was supported by the Research Council of Norway (NFR Grant No.\ 310713).
\end{acknowledgments}


\begin{thebibliography}{40}%
\makeatletter
\providecommand \@ifxundefined [1]{%
 \@ifx{#1\undefined}
}%
\providecommand \@ifnum [1]{%
 \ifnum #1\expandafter \@firstoftwo
 \else \expandafter \@secondoftwo
 \fi
}%
\providecommand \@ifx [1]{%
 \ifx #1\expandafter \@firstoftwo
 \else \expandafter \@secondoftwo
 \fi
}%
\providecommand \natexlab [1]{#1}%
\providecommand \enquote  [1]{``#1''}%
\providecommand \bibnamefont  [1]{#1}%
\providecommand \bibfnamefont [1]{#1}%
\providecommand \citenamefont [1]{#1}%
\providecommand \href@noop [0]{\@secondoftwo}%
\providecommand \href [0]{\begingroup \@sanitize@url \@href}%
\providecommand \@href[1]{\@@startlink{#1}\@@href}%
\providecommand \@@href[1]{\endgroup#1\@@endlink}%
\providecommand \@sanitize@url [0]{\catcode `\\12\catcode `\$12\catcode
  `\&12\catcode `\#12\catcode `\^12\catcode `\_12\catcode `\%12\relax}%
\providecommand \@@startlink[1]{}%
\providecommand \@@endlink[0]{}%
\providecommand \url  [0]{\begingroup\@sanitize@url \@url }%
\providecommand \@url [1]{\endgroup\@href {#1}{\urlprefix }}%
\providecommand \urlprefix  [0]{URL }%
\providecommand \Eprint [0]{\href }%
\providecommand \doibase [0]{https://doi.org/}%
\providecommand \selectlanguage [0]{\@gobble}%
\providecommand \bibinfo  [0]{\@secondoftwo}%
\providecommand \bibfield  [0]{\@secondoftwo}%
\providecommand \translation [1]{[#1]}%
\providecommand \BibitemOpen [0]{}%
\providecommand \bibitemStop [0]{}%
\providecommand \bibitemNoStop [0]{.\EOS\space}%
\providecommand \EOS [0]{\spacefactor3000\relax}%
\providecommand \BibitemShut  [1]{\csname bibitem#1\endcsname}%
\let\auto@bib@innerbib\@empty
\bibitem [{\citenamefont {Peauger}\ \emph {et~al.}(2011)\citenamefont
  {Peauger}, \citenamefont {Farabolini}, \citenamefont {Girardot},
  \citenamefont {Andersson}, \citenamefont {Riddone}, \citenamefont
  {Samoshkin}, \citenamefont {Solodko}, \citenamefont {Zennaro},\ and\
  \citenamefont {Ruber}}]{peauger_wakefield_2011}%
  \BibitemOpen
  \bibfield  {author} {\bibinfo {author} {\bibfnamefont {F.}~\bibnamefont
  {Peauger}}, \bibinfo {author} {\bibfnamefont {W.}~\bibnamefont {Farabolini}},
  \bibinfo {author} {\bibfnamefont {P.}~\bibnamefont {Girardot}}, \bibinfo
  {author} {\bibfnamefont {A.}~\bibnamefont {Andersson}}, \bibinfo {author}
  {\bibfnamefont {G.}~\bibnamefont {Riddone}}, \bibinfo {author} {\bibfnamefont
  {A.}~\bibnamefont {Samoshkin}}, \bibinfo {author} {\bibfnamefont
  {A.}~\bibnamefont {Solodko}}, \bibinfo {author} {\bibfnamefont
  {R.}~\bibnamefont {Zennaro}},\ and\ \bibinfo {author} {\bibfnamefont
  {R.}~\bibnamefont {Ruber}},\ }\bibfield  {title} {\bibinfo {title}
  {{WAKEFIELD} {MONITOR} {DEVELOPMENT} {FOR} {CLIC} {ACCELERATING}
  {STRUCTURE}},\ }in\ \href@noop {} {\emph {\bibinfo {booktitle} {25th
  {International} {Linear} {Accelerator} {Conference}}}}\ (\bibinfo {year}
  {2011})\ p.~\bibinfo {pages} {3}\BibitemShut {NoStop}%
\bibitem [{\citenamefont {{CLIC
  collaboration}}(2018)}]{clic_collaboration_compact_2018}%
  \BibitemOpen
  \bibfield  {author} {\bibinfo {author} {\bibnamefont {{CLIC
  collaboration}}},\ }\href {https://doi.org/10.23731/CYRM-2018-004} {\emph
  {\bibinfo {title} {The {Compact} {Linear} {Collider} ({CLIC}) – {Project}
  {Implementation} {Plan}}}},\ \bibinfo {type} {Tech. Rep.}\ (\bibinfo
  {institution} {CERN},\ \bibinfo {year} {2018})\BibitemShut {NoStop}%
\bibitem [{\citenamefont {{CLIC
  collaboration}}(2016)}]{clic_collaboration_updated_2016}%
  \BibitemOpen
  \bibfield  {author} {\bibinfo {author} {\bibnamefont {{CLIC
  collaboration}}},\ }\href {https://doi.org/10.5170/CERN-2016-004} {\emph
  {\bibinfo {title} {Updated baseline for a staged {Compact} {Linear}
  {Collider}}}}\ (\bibinfo  {publisher} {CERN},\ \bibinfo {year}
  {2016})\BibitemShut {NoStop}%
\bibitem [{\citenamefont {{CLIC
  collaboration}}(2012)}]{clic_collaboration_multi-tev_2012}%
  \BibitemOpen
  \bibfield  {author} {\bibinfo {author} {\bibnamefont {{CLIC
  collaboration}}},\ }\href {https://doi.org/10.5170/CERN-2012-007} {\emph
  {\bibinfo {title} {A {Multi}-{TeV} {Linear} {Collider} {Based} on {CLIC}
  {Technology}: {CLIC} {Conceptual} {Design} {Report}}}}\ (\bibinfo
  {publisher} {CERN},\ \bibinfo {year} {2012})\BibitemShut {NoStop}%
\bibitem [{\citenamefont {Kraljevic}\ and\ \citenamefont
  {Schulte}(2019)}]{kraljevic_beam-based_2019}%
  \BibitemOpen
  \bibfield  {author} {\bibinfo {author} {\bibfnamefont {N.~B.}\ \bibnamefont
  {Kraljevic}}\ and\ \bibinfo {author} {\bibfnamefont {D.}~\bibnamefont
  {Schulte}},\ }\bibfield  {title} {\bibinfo {title} {Beam-{Based} {Beamline}
  {Element} {Alignment} for the {Main} {Linac} of the 380 {GeV} {Stage} of
  {CLIC}}\ }(\bibinfo {year} {2019})\ p.~\bibinfo {pages} {3}\BibitemShut
  {NoStop}%
\bibitem [{\citenamefont {Gilardi}(2021)}]{gilardi_measurements_2021}%
  \BibitemOpen
  \bibfield  {author} {\bibinfo {author} {\bibfnamefont {A.}~\bibnamefont
  {Gilardi}},\ }\emph {\bibinfo {title} {Measurements of wakefields and bunch
  length with beam in linear electron accelerators: a case study at the {CLEAR}
  facility}},\ \href {https://cds.cern.ch/record/2766988/} {\bibinfo {type}
  {{PhD}}},\ \bibinfo  {school} {University of Naples Federico II}, \bibinfo
  {address} {Naples, Italy} (\bibinfo {year} {2021})\BibitemShut {NoStop}%
\bibitem [{\citenamefont {{Lillestøl, Reidar
  Lunde}}(2017)}]{lillestol_reidar_lunde_clic_2017}%
  \BibitemOpen
  \bibfield  {author} {\bibinfo {author} {\bibnamefont {{Lillestøl, Reidar
  Lunde}}},\ }\href {https://cds.cern.ch/record/2751094} {\emph {\bibinfo
  {title} {The {CLIC} {Wake} {Field} {Monitor} {Study}}}},\ \bibinfo {type}
  {Tech. Rep.}\ \bibinfo {number} {1169}\ (\bibinfo  {institution} {CERN},\
  \bibinfo {year} {2017})\BibitemShut {NoStop}%
\bibitem [{\citenamefont {Lillestol}\ \emph {et~al.}(2016)\citenamefont
  {Lillestol}, \citenamefont {Lillestøl}, \citenamefont {Adli},\ and\
  \citenamefont {Pfingstner}}]{lillestol_status_2016}%
  \BibitemOpen
  \bibfield  {author} {\bibinfo {author} {\bibfnamefont {R.~L.}\ \bibnamefont
  {Lillestol}}, \bibinfo {author} {\bibfnamefont {R.~L.}\ \bibnamefont
  {Lillestøl}}, \bibinfo {author} {\bibfnamefont {E.}~\bibnamefont {Adli}},\
  and\ \bibinfo {author} {\bibfnamefont {J.}~\bibnamefont {Pfingstner}},\
  }\bibfield  {title} {\bibinfo {title} {Status of {Wakefield} {Monitor}
  {Experiments} at the {CLIC} {Test} {Facility}}\ }(\bibinfo {year} {2016})\
  p.~\bibinfo {pages} {3}\BibitemShut {NoStop}%
\bibitem [{\citenamefont {Lillestol}\ \emph {et~al.}(2015)\citenamefont
  {Lillestol}, \citenamefont {Adli}, \citenamefont {Pfingstner}, \citenamefont
  {Farabolini}, \citenamefont {Corsini}, \citenamefont {Döbert}, \citenamefont
  {Grudiev}, \citenamefont {Malina},\ and\ \citenamefont
  {Wuensch}}]{lillestol_wakefield_2015}%
  \BibitemOpen
  \bibfield  {author} {\bibinfo {author} {\bibfnamefont {R.~L.}\ \bibnamefont
  {Lillestol}}, \bibinfo {author} {\bibfnamefont {E.}~\bibnamefont {Adli}},
  \bibinfo {author} {\bibfnamefont {J.}~\bibnamefont {Pfingstner}}, \bibinfo
  {author} {\bibfnamefont {W.}~\bibnamefont {Farabolini}}, \bibinfo {author}
  {\bibfnamefont {R.}~\bibnamefont {Corsini}}, \bibinfo {author} {\bibfnamefont
  {S.}~\bibnamefont {Döbert}}, \bibinfo {author} {\bibfnamefont
  {A.}~\bibnamefont {Grudiev}}, \bibinfo {author} {\bibfnamefont
  {L.}~\bibnamefont {Malina}},\ and\ \bibinfo {author} {\bibfnamefont
  {W.}~\bibnamefont {Wuensch}},\ }\bibfield  {title} {\bibinfo {title}
  {Wakefield {Monitor} {Experiments} with {X}-{Band} {Accelerating}
  {Structures}}\ }(\bibinfo {year} {2015})\ p.~\bibinfo {pages} {3}\BibitemShut
  {NoStop}%
\bibitem [{\citenamefont {Quirante}\ \emph {et~al.}(2014)\citenamefont
  {Quirante}, \citenamefont {Corsini}, \citenamefont {Farabolini},
  \citenamefont {Gamba}, \citenamefont {Grudiev}, \citenamefont {Khan},
  \citenamefont {Lefevre}, \citenamefont {Lefèvre}, \citenamefont {Mazzoni},
  \citenamefont {Pan}, \citenamefont {Peauger}, \citenamefont {Tecker},
  \citenamefont {Vitoratou},\ and\ \citenamefont
  {Yaqub}}]{quirante_califes_2014}%
  \BibitemOpen
  \bibfield  {author} {\bibinfo {author} {\bibfnamefont {J.~L.~N.}\
  \bibnamefont {Quirante}}, \bibinfo {author} {\bibfnamefont {R.}~\bibnamefont
  {Corsini}}, \bibinfo {author} {\bibfnamefont {W.}~\bibnamefont {Farabolini}},
  \bibinfo {author} {\bibfnamefont {D.}~\bibnamefont {Gamba}}, \bibinfo
  {author} {\bibfnamefont {A.}~\bibnamefont {Grudiev}}, \bibinfo {author}
  {\bibfnamefont {M.~A.}\ \bibnamefont {Khan}}, \bibinfo {author}
  {\bibfnamefont {T.}~\bibnamefont {Lefevre}}, \bibinfo {author} {\bibfnamefont
  {T.}~\bibnamefont {Lefèvre}}, \bibinfo {author} {\bibfnamefont
  {S.}~\bibnamefont {Mazzoni}}, \bibinfo {author} {\bibfnamefont
  {R.}~\bibnamefont {Pan}}, \bibinfo {author} {\bibfnamefont {F.}~\bibnamefont
  {Peauger}}, \bibinfo {author} {\bibfnamefont {F.}~\bibnamefont {Tecker}},
  \bibinfo {author} {\bibfnamefont {N.}~\bibnamefont {Vitoratou}},\ and\
  \bibinfo {author} {\bibfnamefont {K.}~\bibnamefont {Yaqub}},\ }\bibfield
  {title} {\bibinfo {title} {{CALIFES}: {A} {Multi}-{Purpose} {Electron} {Beam}
  for {Accelerator} {Technology} {Tests}}\ }(\bibinfo  {publisher} {JACOW,
  Geneva, Switzerland},\ \bibinfo {address} {Geneva, Switzerland},\ \bibinfo
  {year} {2014})\ p.~\bibinfo {pages} {3}\BibitemShut {NoStop}%
\bibitem [{\citenamefont {Farabolini}\ \emph {et~al.}(2014)\citenamefont
  {Farabolini}, \citenamefont {Borgmann}, \citenamefont {Corsini},
  \citenamefont {Gamba}, \citenamefont {Grudiev}, \citenamefont {Khan},
  \citenamefont {Mazzoni}, \citenamefont {Navarro~Quirante}, \citenamefont
  {Ögren}, \citenamefont {Pan}, \citenamefont {Peauger}, \citenamefont
  {Ruber}, \citenamefont {Towler}, \citenamefont {Vitoratou},\ and\
  \citenamefont {Yaqub}}]{farabolini_recent_2014}%
  \BibitemOpen
  \bibfield  {author} {\bibinfo {author} {\bibfnamefont {W.}~\bibnamefont
  {Farabolini}}, \bibinfo {author} {\bibfnamefont {C.}~\bibnamefont
  {Borgmann}}, \bibinfo {author} {\bibfnamefont {R.}~\bibnamefont {Corsini}},
  \bibinfo {author} {\bibfnamefont {D.}~\bibnamefont {Gamba}}, \bibinfo
  {author} {\bibfnamefont {A.}~\bibnamefont {Grudiev}}, \bibinfo {author}
  {\bibfnamefont {M.~A.}\ \bibnamefont {Khan}}, \bibinfo {author}
  {\bibfnamefont {S.}~\bibnamefont {Mazzoni}}, \bibinfo {author} {\bibfnamefont
  {J.~L.}\ \bibnamefont {Navarro~Quirante}}, \bibinfo {author} {\bibfnamefont
  {J.}~\bibnamefont {Ögren}}, \bibinfo {author} {\bibfnamefont
  {R.}~\bibnamefont {Pan}}, \bibinfo {author} {\bibfnamefont {F.}~\bibnamefont
  {Peauger}}, \bibinfo {author} {\bibfnamefont {R.}~\bibnamefont {Ruber}},
  \bibinfo {author} {\bibfnamefont {J.}~\bibnamefont {Towler}}, \bibinfo
  {author} {\bibfnamefont {N.}~\bibnamefont {Vitoratou}},\ and\ \bibinfo
  {author} {\bibfnamefont {K.}~\bibnamefont {Yaqub}},\ }\bibfield  {title}
  {\bibinfo {title} {Recent {Results} from {CTF3} {Two} {Beam} {Test} {Stand}}\
  }(\bibinfo  {publisher} {JACOW Publishing, Geneva, Switzerland},\ \bibinfo
  {year} {2014})\ pp.\ \bibinfo {pages} {1880--1882}\BibitemShut {NoStop}%
\bibitem [{\citenamefont {Gamba}\ \emph {et~al.}(2017)\citenamefont {Gamba},
  \citenamefont {Corsini}, \citenamefont {Curt}, \citenamefont {Doebert},
  \citenamefont {Farabolini}, \citenamefont {Mcmonagle}, \citenamefont
  {Skowronski}, \citenamefont {Tecker}, \citenamefont {Zeeshan}, \citenamefont
  {Adli}, \citenamefont {Lindstrøm}, \citenamefont {Ross},\ and\ \citenamefont
  {Wroe}}]{gamba_clear_2017}%
  \BibitemOpen
  \bibfield  {author} {\bibinfo {author} {\bibfnamefont {D.}~\bibnamefont
  {Gamba}}, \bibinfo {author} {\bibfnamefont {R.}~\bibnamefont {Corsini}},
  \bibinfo {author} {\bibfnamefont {S.}~\bibnamefont {Curt}}, \bibinfo {author}
  {\bibfnamefont {S.}~\bibnamefont {Doebert}}, \bibinfo {author} {\bibfnamefont
  {W.}~\bibnamefont {Farabolini}}, \bibinfo {author} {\bibfnamefont
  {G.}~\bibnamefont {Mcmonagle}}, \bibinfo {author} {\bibfnamefont
  {P.}~\bibnamefont {Skowronski}}, \bibinfo {author} {\bibfnamefont
  {F.}~\bibnamefont {Tecker}}, \bibinfo {author} {\bibfnamefont
  {S.}~\bibnamefont {Zeeshan}}, \bibinfo {author} {\bibfnamefont
  {E.}~\bibnamefont {Adli}}, \bibinfo {author} {\bibfnamefont {C.}~\bibnamefont
  {Lindstrøm}}, \bibinfo {author} {\bibfnamefont {A.}~\bibnamefont {Ross}},\
  and\ \bibinfo {author} {\bibfnamefont {L.}~\bibnamefont {Wroe}},\ }\bibfield
  {title} {\bibinfo {title} {The {CLEAR} user facility at {CERN}},\ }\bibfield
  {journal} {\bibinfo  {journal} {Nuclear Instruments and Methods in Physics
  Research Section A: Accelerators, Spectrometers, Detectors and Associated
  Equipment}\ }\href {https://doi.org/10.1016/j.nima.2017.11.080}
  {10.1016/j.nima.2017.11.080} (\bibinfo {year} {2017})\BibitemShut {NoStop}%
\bibitem [{\citenamefont {Sjobak}\ \emph {et~al.}(2019)\citenamefont {Sjobak},
  \citenamefont {Adli}, \citenamefont {Bergamaschi}, \citenamefont {Burger},
  \citenamefont {Corsini}, \citenamefont {Curcio}, \citenamefont {Curt},
  \citenamefont {Döbert}, \citenamefont {Farabolini}, \citenamefont {Gamba},
  \citenamefont {Garolfi}, \citenamefont {Gilardi}, \citenamefont {Gorgisyan},
  \citenamefont {Granados}, \citenamefont {Guerin}, \citenamefont {Kieffer},
  \citenamefont {Krupa}, \citenamefont {Lefèvre}, \citenamefont {Lindstrøm},
  \citenamefont {Lyapin}, \citenamefont {Mazzoni}, \citenamefont {McMonagle},
  \citenamefont {Nadenau}, \citenamefont {Panuganti}, \citenamefont {Pitman},
  \citenamefont {Rude}, \citenamefont {Schlogelhofer}, \citenamefont
  {Skowroński}, \citenamefont {Wendt},\ and\ \citenamefont
  {Zemanek}}]{sjobak_status_2019}%
  \BibitemOpen
  \bibfield  {author} {\bibinfo {author} {\bibfnamefont {K.}~\bibnamefont
  {Sjobak}}, \bibinfo {author} {\bibfnamefont {E.}~\bibnamefont {Adli}},
  \bibinfo {author} {\bibfnamefont {M.}~\bibnamefont {Bergamaschi}}, \bibinfo
  {author} {\bibfnamefont {S.}~\bibnamefont {Burger}}, \bibinfo {author}
  {\bibfnamefont {R.}~\bibnamefont {Corsini}}, \bibinfo {author} {\bibfnamefont
  {A.}~\bibnamefont {Curcio}}, \bibinfo {author} {\bibfnamefont
  {S.}~\bibnamefont {Curt}}, \bibinfo {author} {\bibfnamefont {S.}~\bibnamefont
  {Döbert}}, \bibinfo {author} {\bibfnamefont {W.}~\bibnamefont {Farabolini}},
  \bibinfo {author} {\bibfnamefont {D.}~\bibnamefont {Gamba}}, \bibinfo
  {author} {\bibfnamefont {L.}~\bibnamefont {Garolfi}}, \bibinfo {author}
  {\bibfnamefont {A.}~\bibnamefont {Gilardi}}, \bibinfo {author} {\bibfnamefont
  {I.}~\bibnamefont {Gorgisyan}}, \bibinfo {author} {\bibfnamefont
  {E.}~\bibnamefont {Granados}}, \bibinfo {author} {\bibfnamefont
  {H.}~\bibnamefont {Guerin}}, \bibinfo {author} {\bibfnamefont
  {R.}~\bibnamefont {Kieffer}}, \bibinfo {author} {\bibfnamefont
  {M.}~\bibnamefont {Krupa}}, \bibinfo {author} {\bibfnamefont
  {T.}~\bibnamefont {Lefèvre}}, \bibinfo {author} {\bibfnamefont
  {C.}~\bibnamefont {Lindstrøm}}, \bibinfo {author} {\bibfnamefont
  {A.}~\bibnamefont {Lyapin}}, \bibinfo {author} {\bibfnamefont
  {S.}~\bibnamefont {Mazzoni}}, \bibinfo {author} {\bibfnamefont
  {G.}~\bibnamefont {McMonagle}}, \bibinfo {author} {\bibfnamefont
  {J.}~\bibnamefont {Nadenau}}, \bibinfo {author} {\bibfnamefont
  {H.}~\bibnamefont {Panuganti}}, \bibinfo {author} {\bibfnamefont
  {S.}~\bibnamefont {Pitman}}, \bibinfo {author} {\bibfnamefont
  {V.}~\bibnamefont {Rude}}, \bibinfo {author} {\bibfnamefont {A.}~\bibnamefont
  {Schlogelhofer}}, \bibinfo {author} {\bibfnamefont {P.}~\bibnamefont
  {Skowroński}}, \bibinfo {author} {\bibfnamefont {M.}~\bibnamefont {Wendt}},\
  and\ \bibinfo {author} {\bibfnamefont {A.}~\bibnamefont {Zemanek}},\
  }\bibfield  {title} {\bibinfo {title} {Status of the {CLEAR} {Electron}
  {Beam} {User} {Facility} at {CERN}},\ }in\ \href
  {https://doi.org/10.18429/jacow-ipac2019-mopts054} {\emph {\bibinfo
  {booktitle} {Proceedings of the 10th {Int}. {Partile} {Accelerator} {Conf}.
  {IPAC2019}}}}\ (\bibinfo {year} {2019})\BibitemShut {NoStop}%
\bibitem [{\citenamefont {Dehler}\ \emph
  {et~al.}(2009{\natexlab{a}})\citenamefont {Dehler}, \citenamefont {Raguin},
  \citenamefont {Citterio}, \citenamefont {Falone}, \citenamefont {Wuensch},
  \citenamefont {Riddone}, \citenamefont {Grudiev},\ and\ \citenamefont
  {Zennaro}}]{dehler_x-band_2009}%
  \BibitemOpen
  \bibfield  {author} {\bibinfo {author} {\bibfnamefont {M.}~\bibnamefont
  {Dehler}}, \bibinfo {author} {\bibfnamefont {J.-Y.}\ \bibnamefont {Raguin}},
  \bibinfo {author} {\bibfnamefont {A.}~\bibnamefont {Citterio}}, \bibinfo
  {author} {\bibfnamefont {A.}~\bibnamefont {Falone}}, \bibinfo {author}
  {\bibfnamefont {W.}~\bibnamefont {Wuensch}}, \bibinfo {author} {\bibfnamefont
  {G.}~\bibnamefont {Riddone}}, \bibinfo {author} {\bibfnamefont
  {A.}~\bibnamefont {Grudiev}},\ and\ \bibinfo {author} {\bibfnamefont
  {R.}~\bibnamefont {Zennaro}},\ }\bibfield  {title} {\bibinfo {title} {X-band
  rf structure with integrated alignment monitors},\ }\href
  {https://doi.org/10.1103/PhysRevSTAB.12.062001} {\bibfield  {journal}
  {\bibinfo  {journal} {Physical Review Special Topics - Accelerators and
  Beams}\ }\textbf {\bibinfo {volume} {12}},\ \bibinfo {pages} {062001}
  (\bibinfo {year} {2009}{\natexlab{a}})},\ \bibinfo {note} {publisher:
  American Physical Society}\BibitemShut {NoStop}%
\bibitem [{\citenamefont {Dehler}\ \emph
  {et~al.}(2009{\natexlab{b}})\citenamefont {Dehler}, \citenamefont {Raguin},
  \citenamefont {Citterio}, \citenamefont {Falone}, \citenamefont {Wuensch},
  \citenamefont {Riddone}, \citenamefont {Grudiev}, \citenamefont {Zennaro},
  \citenamefont {Samoshkin}, \citenamefont {Gudkov}, \citenamefont {D'Auria},\
  and\ \citenamefont {Elashmawy}}]{dehler_alignment_2009}%
  \BibitemOpen
  \bibfield  {author} {\bibinfo {author} {\bibfnamefont {M.~M.}\ \bibnamefont
  {Dehler}}, \bibinfo {author} {\bibfnamefont {J.-Y.}\ \bibnamefont {Raguin}},
  \bibinfo {author} {\bibfnamefont {A.}~\bibnamefont {Citterio}}, \bibinfo
  {author} {\bibfnamefont {A.}~\bibnamefont {Falone}}, \bibinfo {author}
  {\bibfnamefont {W.}~\bibnamefont {Wuensch}}, \bibinfo {author} {\bibfnamefont
  {G.}~\bibnamefont {Riddone}}, \bibinfo {author} {\bibfnamefont
  {A.}~\bibnamefont {Grudiev}}, \bibinfo {author} {\bibfnamefont
  {R.}~\bibnamefont {Zennaro}}, \bibinfo {author} {\bibfnamefont
  {A.}~\bibnamefont {Samoshkin}}, \bibinfo {author} {\bibfnamefont
  {D.}~\bibnamefont {Gudkov}}, \bibinfo {author} {\bibfnamefont
  {G.}~\bibnamefont {D'Auria}},\ and\ \bibinfo {author} {\bibfnamefont
  {M.}~\bibnamefont {Elashmawy}},\ }\bibfield  {title} {\bibinfo {title}
  {Alignment {Monitors} for an {X} {Band} {Accelerating} {Structure}}\
  }(\bibinfo {address} {Basel, Switzerland},\ \bibinfo {year} {2009})\
  p.~\bibinfo {pages} {3}\BibitemShut {NoStop}%
\bibitem [{\citenamefont {Dehler}\ \emph {et~al.}(2013)\citenamefont {Dehler},
  \citenamefont {Bettoni}, \citenamefont {Beutner},\ and\ \citenamefont
  {Michele}}]{dehler_wake_2013}%
  \BibitemOpen
  \bibfield  {author} {\bibinfo {author} {\bibfnamefont {M.~M.}\ \bibnamefont
  {Dehler}}, \bibinfo {author} {\bibfnamefont {S.}~\bibnamefont {Bettoni}},
  \bibinfo {author} {\bibfnamefont {B.}~\bibnamefont {Beutner}},\ and\ \bibinfo
  {author} {\bibfnamefont {G.~D.}\ \bibnamefont {Michele}},\ }\bibfield
  {title} {\bibinfo {title} {Wake {Field} {Monitors} in a {Multi} {Purpose} {X}
  {Band} {Accelerating} {Structure}}\ }(\bibinfo {year} {2013})\ p.~\bibinfo
  {pages} {4}\BibitemShut {NoStop}%
\bibitem [{\citenamefont {Dehler}\ and\ \citenamefont
  {Hunziker}(2014)}]{dehler_front_2014}%
  \BibitemOpen
  \bibfield  {author} {\bibinfo {author} {\bibfnamefont {M.~M.}\ \bibnamefont
  {Dehler}}\ and\ \bibinfo {author} {\bibfnamefont {S.}~\bibnamefont
  {Hunziker}},\ }\bibfield  {title} {\bibinfo {title} {Front {End} {Concept}
  for a {Wake} {Field} {Monitor}}\ }(\bibinfo {year} {2014})\ p.~\bibinfo
  {pages} {5}\BibitemShut {NoStop}%
\bibitem [{\citenamefont {Seidel}(1997)}]{seidel_studies_1997}%
  \BibitemOpen
  \bibfield  {author} {\bibinfo {author} {\bibfnamefont {M.}~\bibnamefont
  {Seidel}},\ }\bibfield  {title} {\bibinfo {title} {Studies of beam induced
  dipole-mode signals in accelerating structures at the {SLC}},\ }in\ \href
  {https://doi.org/10.1109/PAC.1997.749679} {\emph {\bibinfo {booktitle}
  {Proceedings of the 1997 {Particle} {Accelerator} {Conference} ({Cat}.
  {No}.{97CH36167})}}},\ Vol.~\bibinfo {volume} {1}\ (\bibinfo {year} {1997})\
  pp.\ \bibinfo {pages} {434--438 vol.1}\BibitemShut {NoStop}%
\bibitem [{\citenamefont {Adolphsen}\ \emph {et~al.}(1999)\citenamefont
  {Adolphsen}, \citenamefont {Bane}, \citenamefont {Jones}, \citenamefont
  {Kroll}, \citenamefont {McCormick}, \citenamefont {Miller}, \citenamefont
  {Ross}, \citenamefont {Slaton}, \citenamefont {Wang},\ and\ \citenamefont
  {Higo}}]{adolphsen_wakefield_1999}%
  \BibitemOpen
  \bibfield  {author} {\bibinfo {author} {\bibfnamefont {C.~D.}\ \bibnamefont
  {Adolphsen}}, \bibinfo {author} {\bibfnamefont {K.~L.~F.}\ \bibnamefont
  {Bane}}, \bibinfo {author} {\bibfnamefont {R.~M.}\ \bibnamefont {Jones}},
  \bibinfo {author} {\bibfnamefont {N.~M.}\ \bibnamefont {Kroll}}, \bibinfo
  {author} {\bibfnamefont {D.~J.}\ \bibnamefont {McCormick}}, \bibinfo {author}
  {\bibfnamefont {R.~H.}\ \bibnamefont {Miller}}, \bibinfo {author}
  {\bibfnamefont {M.~C.}\ \bibnamefont {Ross}}, \bibinfo {author}
  {\bibfnamefont {T.}~\bibnamefont {Slaton}}, \bibinfo {author} {\bibfnamefont
  {J.}~\bibnamefont {Wang}},\ and\ \bibinfo {author} {\bibfnamefont
  {T.}~\bibnamefont {Higo}},\ }\bibfield  {title} {\bibinfo {title} {Wakefield
  and {Beam} {Centering} {Measurements} of a {Damped} and {Detuned} {X}-{Band}
  {Accelerator} {Structure}},\ }in\ \href@noop {} {\emph {\bibinfo {booktitle}
  {New {York}}}}\ (\bibinfo {address} {New York, USA},\ \bibinfo {year}
  {1999})\ p.~\bibinfo {pages} {3}\BibitemShut {NoStop}%
\bibitem [{\citenamefont {Dobert}\ \emph {et~al.}(2005)\citenamefont {Dobert},
  \citenamefont {Adolphsen}, \citenamefont {Jones}, \citenamefont
  {Lewandowski}, \citenamefont {Li}, \citenamefont {Pivi}, \citenamefont
  {Wang},\ and\ \citenamefont {Higo}}]{dobert_beam_2005}%
  \BibitemOpen
  \bibfield  {author} {\bibinfo {author} {\bibfnamefont {S.}~\bibnamefont
  {Dobert}}, \bibinfo {author} {\bibfnamefont {C.}~\bibnamefont {Adolphsen}},
  \bibinfo {author} {\bibfnamefont {R.}~\bibnamefont {Jones}}, \bibinfo
  {author} {\bibfnamefont {J.}~\bibnamefont {Lewandowski}}, \bibinfo {author}
  {\bibfnamefont {Z.}~\bibnamefont {Li}}, \bibinfo {author} {\bibfnamefont
  {M.}~\bibnamefont {Pivi}}, \bibinfo {author} {\bibfnamefont {J.}~\bibnamefont
  {Wang}},\ and\ \bibinfo {author} {\bibfnamefont {T.}~\bibnamefont {Higo}},\
  }\bibfield  {title} {\bibinfo {title} {Beam {Position} {Monitoring} {Using}
  the {Hom}-{Signals} from a {Damped} and {Detuned} {Accelerating}
  {Structure}},\ }in\ \href {https://doi.org/10.1109/PAC.2005.1591275} {\emph
  {\bibinfo {booktitle} {Proceedings of the 2005 {Particle} {Accelerator}
  {Conference}}}}\ (\bibinfo  {publisher} {IEEE},\ \bibinfo {address}
  {Knoxville, TN, USA},\ \bibinfo {year} {2005})\ pp.\ \bibinfo {pages}
  {2804--2806}\BibitemShut {NoStop}%
\bibitem [{\citenamefont {Molloy}\ \emph {et~al.}(2007)\citenamefont {Molloy},
  \citenamefont {Frisch}, \citenamefont {McCormick}, \citenamefont {May},
  \citenamefont {Ross}, \citenamefont {Smith}, \citenamefont {Eddy},
  \citenamefont {Nagaitsev}, \citenamefont {Rechenmacher}, \citenamefont
  {Piccoli}, \citenamefont {Baboi}, \citenamefont {Hensler}, \citenamefont
  {Petrosyan}, \citenamefont {Napoly}, \citenamefont {Paparella},\ and\
  \citenamefont {Simon}}]{molloy_high_2007}%
  \BibitemOpen
  \bibfield  {author} {\bibinfo {author} {\bibfnamefont {S.}~\bibnamefont
  {Molloy}}, \bibinfo {author} {\bibfnamefont {J.}~\bibnamefont {Frisch}},
  \bibinfo {author} {\bibfnamefont {D.}~\bibnamefont {McCormick}}, \bibinfo
  {author} {\bibfnamefont {J.}~\bibnamefont {May}}, \bibinfo {author}
  {\bibfnamefont {M.}~\bibnamefont {Ross}}, \bibinfo {author} {\bibfnamefont
  {T.}~\bibnamefont {Smith}}, \bibinfo {author} {\bibfnamefont
  {N.}~\bibnamefont {Eddy}}, \bibinfo {author} {\bibfnamefont {S.}~\bibnamefont
  {Nagaitsev}}, \bibinfo {author} {\bibfnamefont {R.}~\bibnamefont
  {Rechenmacher}}, \bibinfo {author} {\bibfnamefont {L.}~\bibnamefont
  {Piccoli}}, \bibinfo {author} {\bibfnamefont {N.}~\bibnamefont {Baboi}},
  \bibinfo {author} {\bibfnamefont {O.}~\bibnamefont {Hensler}}, \bibinfo
  {author} {\bibfnamefont {L.}~\bibnamefont {Petrosyan}}, \bibinfo {author}
  {\bibfnamefont {O.}~\bibnamefont {Napoly}}, \bibinfo {author} {\bibfnamefont
  {R.}~\bibnamefont {Paparella}},\ and\ \bibinfo {author} {\bibfnamefont
  {C.}~\bibnamefont {Simon}},\ }\bibfield  {title} {\bibinfo {title} {High
  precision {SC} cavity alignment measurements with higher order modes},\
  }\href {https://doi.org/10.1088/0957-0233/18/8/004} {\bibfield  {journal}
  {\bibinfo  {journal} {Measurement Science and Technology}\ }\textbf {\bibinfo
  {volume} {18}},\ \bibinfo {pages} {2314} (\bibinfo {year}
  {2007})}\BibitemShut {NoStop}%
\bibitem [{\citenamefont {Frisch}\ \emph {et~al.}(2006)\citenamefont {Frisch},
  \citenamefont {Hendrickson}, \citenamefont {McCormick}, \citenamefont {May},
  \citenamefont {Molloy}, \citenamefont {Ross}, \citenamefont {Smith},
  \citenamefont {Eddy}, \citenamefont {Nagaitsev}, \citenamefont {Baboi},
  \citenamefont {Hensler}, \citenamefont {Petrosyan}, \citenamefont {Napoly},
  \citenamefont {Paparella},\ and\ \citenamefont {Simon}}]{frisch_high_2006}%
  \BibitemOpen
  \bibfield  {author} {\bibinfo {author} {\bibfnamefont {J.}~\bibnamefont
  {Frisch}}, \bibinfo {author} {\bibfnamefont {L.}~\bibnamefont {Hendrickson}},
  \bibinfo {author} {\bibfnamefont {D.}~\bibnamefont {McCormick}}, \bibinfo
  {author} {\bibfnamefont {J.}~\bibnamefont {May}}, \bibinfo {author}
  {\bibfnamefont {S.}~\bibnamefont {Molloy}}, \bibinfo {author} {\bibfnamefont
  {M.}~\bibnamefont {Ross}}, \bibinfo {author} {\bibfnamefont {T.}~\bibnamefont
  {Smith}}, \bibinfo {author} {\bibfnamefont {N.}~\bibnamefont {Eddy}},
  \bibinfo {author} {\bibfnamefont {S.}~\bibnamefont {Nagaitsev}}, \bibinfo
  {author} {\bibfnamefont {N.}~\bibnamefont {Baboi}}, \bibinfo {author}
  {\bibfnamefont {O.}~\bibnamefont {Hensler}}, \bibinfo {author} {\bibfnamefont
  {L.}~\bibnamefont {Petrosyan}}, \bibinfo {author} {\bibfnamefont
  {O.}~\bibnamefont {Napoly}}, \bibinfo {author} {\bibfnamefont
  {R.}~\bibnamefont {Paparella}},\ and\ \bibinfo {author} {\bibfnamefont
  {C.}~\bibnamefont {Simon}},\ }\bibfield  {title} {\bibinfo {title} {{HIGH}
  {PRECISION} {SC} {CAVITY} {DIAGNOSTICS} {WITH} {HOM} {MEASUREMENTS}}\
  }(\bibinfo {address} {Edinburgh, Scotland},\ \bibinfo {year} {2006})\
  p.~\bibinfo {pages} {5}\BibitemShut {NoStop}%
\bibitem [{\citenamefont {Baboi}\ \emph {et~al.}(2004)\citenamefont {Baboi},
  \citenamefont {Kreps}, \citenamefont {Wendt}, \citenamefont {Devanz},
  \citenamefont {Napoly},\ and\ \citenamefont
  {Paparella}}]{baboi_preliminary_2004}%
  \BibitemOpen
  \bibfield  {author} {\bibinfo {author} {\bibfnamefont {N.}~\bibnamefont
  {Baboi}}, \bibinfo {author} {\bibfnamefont {G.}~\bibnamefont {Kreps}},
  \bibinfo {author} {\bibfnamefont {M.}~\bibnamefont {Wendt}}, \bibinfo
  {author} {\bibfnamefont {G.}~\bibnamefont {Devanz}}, \bibinfo {author}
  {\bibfnamefont {O.}~\bibnamefont {Napoly}},\ and\ \bibinfo {author}
  {\bibfnamefont {R.}~\bibnamefont {Paparella}},\ }\bibfield  {title} {\bibinfo
  {title} {Preliminary {Study} on {HOM}-{Based} {Beam} {Alignment} in the
  {TESLA} {Test} {Facility}}\ }(\bibinfo {address} {Lübeck},\ \bibinfo {year}
  {2004})\ p.~\bibinfo {pages} {3}\BibitemShut {NoStop}%
\bibitem [{\citenamefont {Prochnow}\ \emph {et~al.}(2003)\citenamefont
  {Prochnow}, \citenamefont {Jensen},\ and\ \citenamefont
  {Wuensch}}]{prochnow_measurement_2003}%
  \BibitemOpen
  \bibfield  {author} {\bibinfo {author} {\bibfnamefont {J.}~\bibnamefont
  {Prochnow}}, \bibinfo {author} {\bibfnamefont {E.}~\bibnamefont {Jensen}},\
  and\ \bibinfo {author} {\bibfnamefont {W.}~\bibnamefont {Wuensch}},\
  }\bibfield  {title} {\bibinfo {title} {Measurement of {Beam} {Position}
  {Using} {Highly}-{Damped} {Accelerating} {Structures}}\ }(\bibinfo {address}
  {Portland, Oregon, USA},\ \bibinfo {year} {2003})\ p.~\bibinfo {pages}
  {3}\BibitemShut {NoStop}%
\bibitem [{\citenamefont {Munoz}\ \emph {et~al.}(2015)\citenamefont {Munoz},
  \citenamefont {Catalan-Lasheras}, \citenamefont {Wendt}, \citenamefont
  {Zorzetti},\ and\ \citenamefont {Faus~Golfe}}]{munoz_electromagnetic_2015}%
  \BibitemOpen
  \bibfield  {author} {\bibinfo {author} {\bibfnamefont {N.~G.}\ \bibnamefont
  {Munoz}}, \bibinfo {author} {\bibfnamefont {N.}~\bibnamefont
  {Catalan-Lasheras}}, \bibinfo {author} {\bibfnamefont {M.}~\bibnamefont
  {Wendt}}, \bibinfo {author} {\bibfnamefont {S.}~\bibnamefont {Zorzetti}},\
  and\ \bibinfo {author} {\bibfnamefont {A.}~\bibnamefont {Faus~Golfe}},\
  }\bibfield  {title} {\bibinfo {title} {Electromagnetic {Field}
  {Pre}-alignment of the {Compact} {Linear} {Collider} ({CLIC}) {Accelerating}
  {Structure} with help of {Wakefield} {Monitor} {Signals}}\ }(\bibinfo {year}
  {2015})\ p.~\bibinfo {pages} {5}\BibitemShut {NoStop}%
\bibitem [{\citenamefont {Munoz}\ \emph {et~al.}(2016)\citenamefont {Munoz},
  \citenamefont {Catalan-Lasheras}, \citenamefont {Lasheras},\ and\
  \citenamefont {Grudiev}}]{munoz_pre-alignment_2016}%
  \BibitemOpen
  \bibfield  {author} {\bibinfo {author} {\bibfnamefont {N.~G.}\ \bibnamefont
  {Munoz}}, \bibinfo {author} {\bibfnamefont {N.}~\bibnamefont
  {Catalan-Lasheras}}, \bibinfo {author} {\bibfnamefont {N.~C.}\ \bibnamefont
  {Lasheras}},\ and\ \bibinfo {author} {\bibfnamefont {A.}~\bibnamefont
  {Grudiev}},\ }\bibfield  {title} {\bibinfo {title} {Pre-alignment of
  {Accelerating} {Structures} for {Compact} {Acceleration} and {High}
  {Gradient} using {In}-situ {Radiofrequency} {Methods}}\ }(\bibinfo {year}
  {2016})\ p.~\bibinfo {pages} {4}\BibitemShut {NoStop}%
\bibitem [{\citenamefont {Munoz}\ \emph {et~al.}(2017)\citenamefont {Munoz},
  \citenamefont {Catalan-Lasheras},\ and\ \citenamefont
  {Lasheras}}]{munoz_pre-alignment_2017}%
  \BibitemOpen
  \bibfield  {author} {\bibinfo {author} {\bibfnamefont {N.~G.}\ \bibnamefont
  {Munoz}}, \bibinfo {author} {\bibfnamefont {N.}~\bibnamefont
  {Catalan-Lasheras}},\ and\ \bibinfo {author} {\bibfnamefont {N.~C.}\
  \bibnamefont {Lasheras}},\ }\bibfield  {title} {\bibinfo {title}
  {Pre-{Alignment} {Techniques} {Developments} and {Measurement} {Results} of
  the {Electromagnetic} {Center} of {Warm} {High}-{Gradient} {Accelerating}
  {Structures}}\ }(\bibinfo {year} {2017})\ p.~\bibinfo {pages} {4}\BibitemShut
  {NoStop}%
\bibitem [{\citenamefont {Sjobak}\ \emph {et~al.}(2014)\citenamefont {Sjobak},
  \citenamefont {Grudiev},\ and\ \citenamefont {Adli}}]{sjobak_design_2014}%
  \BibitemOpen
  \bibfield  {author} {\bibinfo {author} {\bibfnamefont {K.~N.}\ \bibnamefont
  {Sjobak}}, \bibinfo {author} {\bibfnamefont {A.}~\bibnamefont {Grudiev}},\
  and\ \bibinfo {author} {\bibfnamefont {E.}~\bibnamefont {Adli}},\ }\href@noop
  {} {\emph {\bibinfo {title} {{DESIGN} {OF} {WAVEGUIDE} {DAMPED} {CELLS} {FOR}
  12 {GHZ} {HIGH} {GRADIENT} {ACCELERATING} {STRUCTURES}}}},\ \bibinfo {type}
  {Tech. Rep.}\ \bibinfo {number} {1026}\ (\bibinfo  {institution} {CERN},\
  \bibinfo {year} {2014})\BibitemShut {NoStop}%
\bibitem [{\citenamefont {De~Michele}\ and\ \citenamefont
  {Grudiev}(2012)}]{de_michele_analysis_2012}%
  \BibitemOpen
  \bibfield  {author} {\bibinfo {author} {\bibfnamefont {G.}~\bibnamefont
  {De~Michele}}\ and\ \bibinfo {author} {\bibfnamefont {A.}~\bibnamefont
  {Grudiev}},\ }\bibfield  {title} {\bibinfo {title} {Analysis of {Long}-range
  {Wakefields} in {CLIC} {Main} {Linac} {Accelerating} {Structures} with
  {Damping} {Loads}},\ }in\ \href@noop {} {\emph {\bibinfo {booktitle}
  {Procedings of {IPAC2012}}}}\ (\bibinfo  {publisher} {JACOW Publishing,
  Geneva, Switzerland},\ \bibinfo {address} {New Orleans, Louisiana, USA},\
  \bibinfo {year} {2012})\ p.~\bibinfo {pages} {3}\BibitemShut {NoStop}%
\bibitem [{\citenamefont {Grudiev}\ and\ \citenamefont
  {Wuensch}(2010)}]{grudiev_design_2010}%
  \BibitemOpen
  \bibfield  {author} {\bibinfo {author} {\bibfnamefont {A.}~\bibnamefont
  {Grudiev}}\ and\ \bibinfo {author} {\bibfnamefont {W.}~\bibnamefont
  {Wuensch}},\ }\bibfield  {title} {\bibinfo {title} {Design of the {CLIC}
  {Main} {Linac} {Accelerating} {Structure} for {CLIC} {Conceptual} {Design}
  {Report}},\ }in\ \href@noop {} {\emph {\bibinfo {booktitle} {Procedings of
  {LINAC2010}}}}\ (\bibinfo {address} {Tsukuba, Japan},\ \bibinfo {year}
  {2010})\ p.~\bibinfo {pages} {3}\BibitemShut {NoStop}%
\bibitem [{\citenamefont {Sjobak}\ and\ \citenamefont
  {Gamba}(2020)}]{sjobak_clear_2020}%
  \BibitemOpen
  \bibfield  {author} {\bibinfo {author} {\bibfnamefont {K.}~\bibnamefont
  {Sjobak}}\ and\ \bibinfo {author} {\bibfnamefont {D.}~\bibnamefont {Gamba}},\
  }\href {https://gitlab.cern.ch/CLEAR/clearlayout/} {\bibinfo {title} {{CLEAR}
  {Layout} {Figures}}} (\bibinfo {year} {2020})\BibitemShut {NoStop}%
\bibitem [{\citenamefont {Arpaia}\ \emph {et~al.}(2020)\citenamefont {Arpaia},
  \citenamefont {Corsini}, \citenamefont {Gilardi}, \citenamefont {Mostacci},
  \citenamefont {Sabato},\ and\ \citenamefont
  {Sjobak}}]{arpaia_enhancing_2020}%
  \BibitemOpen
  \bibfield  {author} {\bibinfo {author} {\bibfnamefont {P.}~\bibnamefont
  {Arpaia}}, \bibinfo {author} {\bibfnamefont {R.}~\bibnamefont {Corsini}},
  \bibinfo {author} {\bibfnamefont {A.}~\bibnamefont {Gilardi}}, \bibinfo
  {author} {\bibfnamefont {A.}~\bibnamefont {Mostacci}}, \bibinfo {author}
  {\bibfnamefont {L.}~\bibnamefont {Sabato}},\ and\ \bibinfo {author}
  {\bibfnamefont {K.~N.}\ \bibnamefont {Sjobak}},\ }\bibfield  {title}
  {\bibinfo {title} {Enhancing particle bunch-length measurements based on
  {Radio} {Frequency} {Deflector} by the use of focusing elements},\ }\href
  {https://doi.org/10.1038/s41598-020-67997-1} {\bibfield  {journal} {\bibinfo
  {journal} {Scientific Reports}\ }\textbf {\bibinfo {volume} {10}},\ \bibinfo
  {pages} {11457} (\bibinfo {year} {2020})},\ \bibinfo {note} {number: 1
  Publisher: Nature Publishing Group}\BibitemShut {NoStop}%
\bibitem [{\citenamefont {Sabato}\ \emph {et~al.}(2021)\citenamefont {Sabato},
  \citenamefont {Arpaia}, \citenamefont {Gilardi}, \citenamefont {Mostacci},
  \citenamefont {Palumbo},\ and\ \citenamefont
  {Variola}}]{sabato_measurement_2021}%
  \BibitemOpen
  \bibfield  {author} {\bibinfo {author} {\bibfnamefont {L.}~\bibnamefont
  {Sabato}}, \bibinfo {author} {\bibfnamefont {P.}~\bibnamefont {Arpaia}},
  \bibinfo {author} {\bibfnamefont {A.}~\bibnamefont {Gilardi}}, \bibinfo
  {author} {\bibfnamefont {A.}~\bibnamefont {Mostacci}}, \bibinfo {author}
  {\bibfnamefont {L.}~\bibnamefont {Palumbo}},\ and\ \bibinfo {author}
  {\bibfnamefont {A.}~\bibnamefont {Variola}},\ }\bibfield  {title} {\bibinfo
  {title} {A {Measurement} {Method} {Based} on {RF} {Deflector} for {Particle}
  {Bunch} {Longitudinal} {Parameters} in {Linear} {Accelerators}},\ }\href
  {https://doi.org/10.1109/TIM.2020.3009342} {\bibfield  {journal} {\bibinfo
  {journal} {IEEE Transactions on Instrumentation and Measurement}\ }\textbf
  {\bibinfo {volume} {70}},\ \bibinfo {pages} {1} (\bibinfo {year} {2021})},\
  \bibinfo {note} {conference Name: IEEE Transactions on Instrumentation and
  Measurement}\BibitemShut {NoStop}%
\bibitem [{\citenamefont {Arpaia}\ \emph {et~al.}(2019)\citenamefont {Arpaia},
  \citenamefont {Corsini}, \citenamefont {Gilardi},\ and\ \citenamefont
  {Sjobak}}]{arpaia_beambased_2019}%
  \BibitemOpen
  \bibfield  {author} {\bibinfo {author} {\bibfnamefont {P.}~\bibnamefont
  {Arpaia}}, \bibinfo {author} {\bibfnamefont {R.}~\bibnamefont {Corsini}},
  \bibinfo {author} {\bibfnamefont {A.}~\bibnamefont {Gilardi}},\ and\ \bibinfo
  {author} {\bibfnamefont {K.~N.}\ \bibnamefont {Sjobak}},\ }\bibfield  {title}
  {\bibinfo {title} {Beam–based alignment of the {CLIC} high-gradient
  {X}-{Band} accelerating structure using beam-screen},\ }in\ \href
  {https://doi.org/10.1109/I2MTC.2019.8827121} {\emph {\bibinfo {booktitle}
  {2019 {IEEE} {International} {Instrumentation} and {Measurement} {Technology}
  {Conference} ({I2MTC})}}}\ (\bibinfo {year} {2019})\ pp.\ \bibinfo {pages}
  {1--6},\ \bibinfo {note} {iSSN: 2642-2077}\BibitemShut {NoStop}%
\bibitem [{\citenamefont {Sjobak}(2018)}]{sjobak_clear_2018}%
  \BibitemOpen
  \bibfield  {author} {\bibinfo {author} {\bibfnamefont {K.~N.}\ \bibnamefont
  {Sjobak}},\ }\href {https://gitlab.cern.ch/CLEAR/CLEARview} {\bibinfo {title}
  {{CLEAR} / {CLEARview}}} (\bibinfo {year} {2018})\BibitemShut {NoStop}%
\bibitem [{\citenamefont {Sjobak}\ \emph {et~al.}(2021)\citenamefont {Sjobak},
  \citenamefont {Korysko}, \citenamefont {Gilardi},\ and\ \citenamefont
  {Farabolini}}]{sjobak_clear_2021}%
  \BibitemOpen
  \bibfield  {author} {\bibinfo {author} {\bibfnamefont {K.~N.}\ \bibnamefont
  {Sjobak}}, \bibinfo {author} {\bibfnamefont {P.}~\bibnamefont {Korysko}},
  \bibinfo {author} {\bibfnamefont {A.}~\bibnamefont {Gilardi}},\ and\ \bibinfo
  {author} {\bibfnamefont {W.}~\bibnamefont {Farabolini}},\ }\href
  {https://gitlab.cern.ch/CLEAR/arduino-positiongaugeserver} {\bibinfo {title}
  {{CLEAR} / {ARDUINO}-{PositionGaugeServer}}} (\bibinfo {year}
  {2021})\BibitemShut {NoStop}%
\bibitem [{\citenamefont {Grudiev}(2011)}]{grudiev_geometry_2011}%
  \BibitemOpen
  \bibfield  {author} {\bibinfo {author} {\bibfnamefont {A.}~\bibnamefont
  {Grudiev}},\ }\href {https://edms.cern.ch/document/1120951/1} {\bibinfo
  {title} {Geometry for simulation of two-beam scheme {\textbar} {Document}
  1120951 (v.1)}} (\bibinfo {year} {2011})\BibitemShut {NoStop}%
\bibitem [{\citenamefont {Solodko}(2012)}]{solodko_rf_2012}%
  \BibitemOpen
  \bibfield  {author} {\bibinfo {author} {\bibfnamefont {A.}~\bibnamefont
  {Solodko}},\ }\href {https://edms.cern.ch/document/1182237/1} {\bibinfo
  {title} {{RF} design of {WFM} for the {CLEX} module accelerating structure
  {TD26}\_cc\_sic {\textbar} {Document} 1182237 (v.1)}} (\bibinfo {year}
  {2012})\BibitemShut {NoStop}%
\bibitem [{\citenamefont {{Alan V. Oppenheim}}\ \emph
  {et~al.}(1996)\citenamefont {{Alan V. Oppenheim}}, \citenamefont {{Alan S.
  Willsky}},\ and\ \citenamefont {{S. Hamid
  Nawab}}}]{alan_v_oppenheim_signals_1996}%
  \BibitemOpen
  \bibfield  {author} {\bibinfo {author} {\bibnamefont {{Alan V. Oppenheim}}},
  \bibinfo {author} {\bibnamefont {{Alan S. Willsky}}},\ and\ \bibinfo {author}
  {\bibnamefont {{S. Hamid Nawab}}},\ }\href@noop {} {\emph {\bibinfo {title}
  {Signals \& {Systems}}}},\ \bibinfo {edition} {2nd}\ ed.\ (\bibinfo
  {publisher} {Prentice Hall},\ \bibinfo {year} {1996})\BibitemShut {NoStop}%
\bibitem [{\citenamefont {{Ronald N.
  Bracewell}}(1999)}]{ronald_n_bracewell_fourier_1999}%
  \BibitemOpen
  \bibfield  {author} {\bibinfo {author} {\bibnamefont {{Ronald N.
  Bracewell}}},\ }\href@noop {} {\emph {\bibinfo {title} {The {Fourier}
  {Transform} \& {It}'s applications}}},\ \bibinfo {edition} {3rd}\ ed.\
  (\bibinfo  {publisher} {McGraw-Hill},\ \bibinfo {year} {1999})\BibitemShut
  {NoStop}%
\end{thebibliography}
%

\end{document}